\documentclass[a4paper]{article}
\usepackage[brl,eqsecnum]{ws1507p}
\makeatletter
\renewcommand{\CreditWS}{%
   Electronic version of an article 
   published as \textit{Biophys.~Rev.~Lett.} \textbf{11}, 9--38 (2016), 
   \allowbreak  
   DOI:~10.1142/S1793048015400019 \space
   \copyright~{World Scientific Publishing Company}\space
   \allowbreak
   \texttt{\@Journal@URL@WS}%
}%
\makeatother
\usepackage[super]{cite}
\def\refcite{\citen}
\usepackage{amsmath}
\usepackage{amsfonts}
\usepackage{bm}
\usepackage{revsymb}
\usepackage{macro1507o}
\usepackage{graphicx}
\graphicspath{{.}}
\newcommand{\IncFig}[2][clip]{\includegraphics[#1]{#2}}
\newcommand{\Uex}{U_{\text{ex}}}
\newcommand{\tauHop}{\tau_{\text{hop}}}
\DeclareMathOperator{\erfc}{erfc}
\begin{document}
\markboth{Ooshida T., Goto S., Matsumoto T., \& Otsuki M.}%
{Insights from SFD into Cooperativity in Higher Dimensions}

\catchline{}{}{}{}{}

\title{Insights from Single-File Diffusion \\
into Co\-operativity in Higher Dimensions}

\providecommand{\surname}{\textsc}
\author{%
   \surname{Ooshida}, Takeshi\thanks{%
   Dept.~Mechanical \& Aerospace Engineering,
   Tottori University, 
   Tottori 680-8552, Japan}~;
   \surname{Goto}, Susumu\thanks{%
   Graduate School of Engineering Science,  
   Osaka University, Toyonaka,
   Osaka 560-8531, Japan}~;
   \\
   \surname{Matsumoto}, Takeshi\thanks{%
   Div.~Physics \& Astronomy, 
   Graduate School of Science, 
   Kyoto University, 
   Kyoto 606-8502, Japan}~;
   \surname{Otsuki}, Michio\thanks{%
   Dept.~Materials Science, 
   Shimane University,
   Matsue 690-8504, Japan}
}

\maketitle

\begin{history}
  \received{Day Month Year}
  \revised{Day Month Year}
\end{history}

\begin{abstract}%
  Diffusion in colloidal suspensions can be very slow 
  due to the cage effect, 
  which confines each particle within a short radius on one hand,  
  and involves large-scale cooperative motions on the other.
  In search of insight into this cooperativity, 
  here the authors develop a formalism 
  to calculate the displacement correlation 
  in colloidal systems, 
  mainly in the two-dimensional case.
  To clarify the idea for it, 
  studies are reviewed
  on cooperativity among the particles
  in the one-dimensional case, 
  i.e.\ 
  the single-file diffusion (SFD).  
  As an improvement over the celebrated formula 
  by Alexander and Pincus 
  on the mean-square displacement (MSD) in SFD,  
  it is shown 
  that the displacement correlation in SFD 
  can be calculated 
  from Lagrangian correlation 
  of the particle interval in the one-dimensional case, 
  and also 
  that the formula can be extended to higher dimensions.
  The improved formula becomes exact for large systems.  
  By combining the formula 
  with a nonlinear theory for correlation, 
  a correction 
  to the asymptotic law for the MSD in SFD 
  is obtained.
  In the two-dimensional case, 
  the linear theory 
  gives description of vortical cooperative motion.
  \vspace{0.5\baselineskip}%
  \par\noindent\relax
  \textit{Special Issue Comments}:
  This article presents methodological insights 
  into mathematical treatment of particle cooperativity 
  in colloidal liquids. 
  This article is related to the Special Issue articles  
  [in Biophys.~Rev.~Lett.]
  about mathematical approaches to single-file diffusion%
  \cite{Taloni.BRL9}
  and Fourier-based analysis of quasi-one-dimensional systems%
  \cite{Coste.BRL9}.  
  \vspace{0.5\baselineskip}%
  %
\end{abstract}

\section{Introduction}
\label{sec:intro}

The slowdown of diffusion 
  in systems of densely packed particles,
  as well as constrained dynamics in many other systems 
  described as ``glassy'', ``jammed'', or ``crowded'',
  has attracted attention\cite{Liu.Book2001}, 
  not only in connection 
  with the physics 
  of glass transition\cite{Berthier.RMP83,Das.Book2011} 
  but also in the context 
  of anomalous transport in biological cells\cite{Hoefling.RPP76}.
The slowdown
  is due to suppression of free diffusion, 
  which implies that the particles can diffuse  
  only in some cooperative way
  with space-time correlations%
  \cite{Yamamoto.PRE58,Berthier.Book2011}.

For concreteness, 
  we consider a colloidal system 
  consisting of spherical Brownian particles
  in the $\nd$-dimensional space, 
  with the representative diameter $\sigma$.
We denote 
  the position vector of the $i$-th particle 
  with $\mb{r}_i(t)$;
  there are $N$ particles so that $i=1,2,\ldots,N$.
The particles 
  are governed by the Langevin equation,  
\begin{equation}
  m \ddot{\rvect}_i
  = -\mu \dot{\rvect}_i  
    - \dd{}{\rvect_i} \sum_{j<k} V_{jk}
    + \mu \mb{f}_i(t)
  \label{Langevin.rvect}, 
\end{equation}
  where 
  $m$ is the mass of the particle, 
  $V_{jk}$ is the interaction potential 
  between the $j$-th and $k$-th particles,
  and
  $\mu\mb{f}_i(t)$ is the random force 
  characterized by the variance
\begin{equation}
  \Av{\mb{f}_i(t) \otimes \mb{f}_j(t')}
  = \frac{2\kT}{\mu} \delta_{ij} \delta(t-t') \openone
  \label{fvect}.
\end{equation}%
The drag coefficient $\mu$, 
  which may depend on the configuration 
  of the particles 
  in general,
  is regarded here as constant for simplicity. 
The system is assumed 
  to fill a periodic box of the linear dimension $L$
  and to be uniform statistically,
  unless specified otherwise, 
  and we consider the thermodynamic limit, $L\to\infty$,  
  keeping the mean density $\rho_0 = N/L^\nd$ fixed.

To study slowdown of diffusion, 
  let us start with denoting 
  the displacement of the $j$-th particle 
  with $\mb{R}_j(t,s) = \mb{r}_j(t) - \mb{r}_j(s)$, 
  where $s < t$; 
  if the system is statistically steady,
  it suffices to consider 
  $\mb{R}_j(t) = \mb{r}_j(t) - \mb{r}_j(0)$.
The mean-square displacement (MSD) 
  is given by $\Av{\mb{R}^2} = \Av{[\mb{R}_j(t)]^2}$,  
  with $\Av{~\cdot~}$ denoting the ensemble average 
  in regard to the random force, as in Eq.~(\ref{fvect}), 
  and the subscript $j$ is dropped 
  because the system is uniform. 
In the case of free diffusion, 
  the MSD grows basically as 
\begin{equation}
  \Av{\mb{R}^2} = 2{\nd}Dt 
  \quad (\text{free diffusion})
  \label{MSD.free},
\end{equation}  
  except for very short inertial time\-scale,
  where $D = \kT/\mu$ is the diffusion constant.

\begin{figure}
  \centering
  \raisebox{3.5cm}{(a)}\hspace*{-0.75em}%
  \IncFig[clip,width=0.35\linewidth]{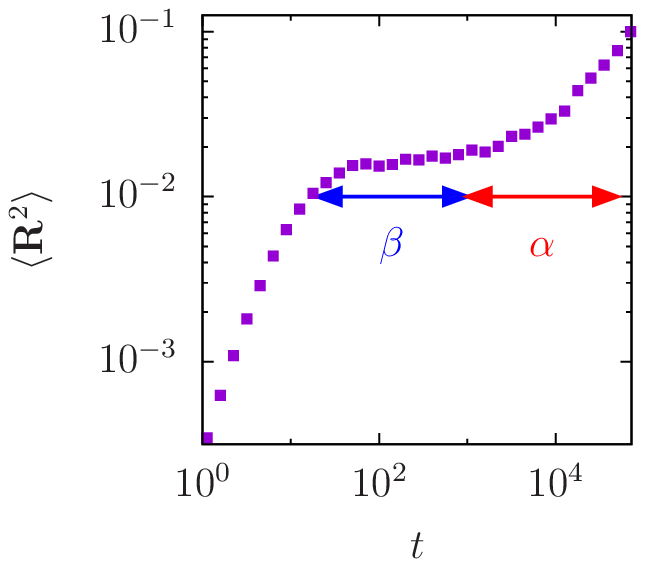}
  \quad
  \raisebox{3.5cm}{(b)}\hspace*{-0.75em}%
  \smash{%
  \IncFig[clip,width=0.47\linewidth]{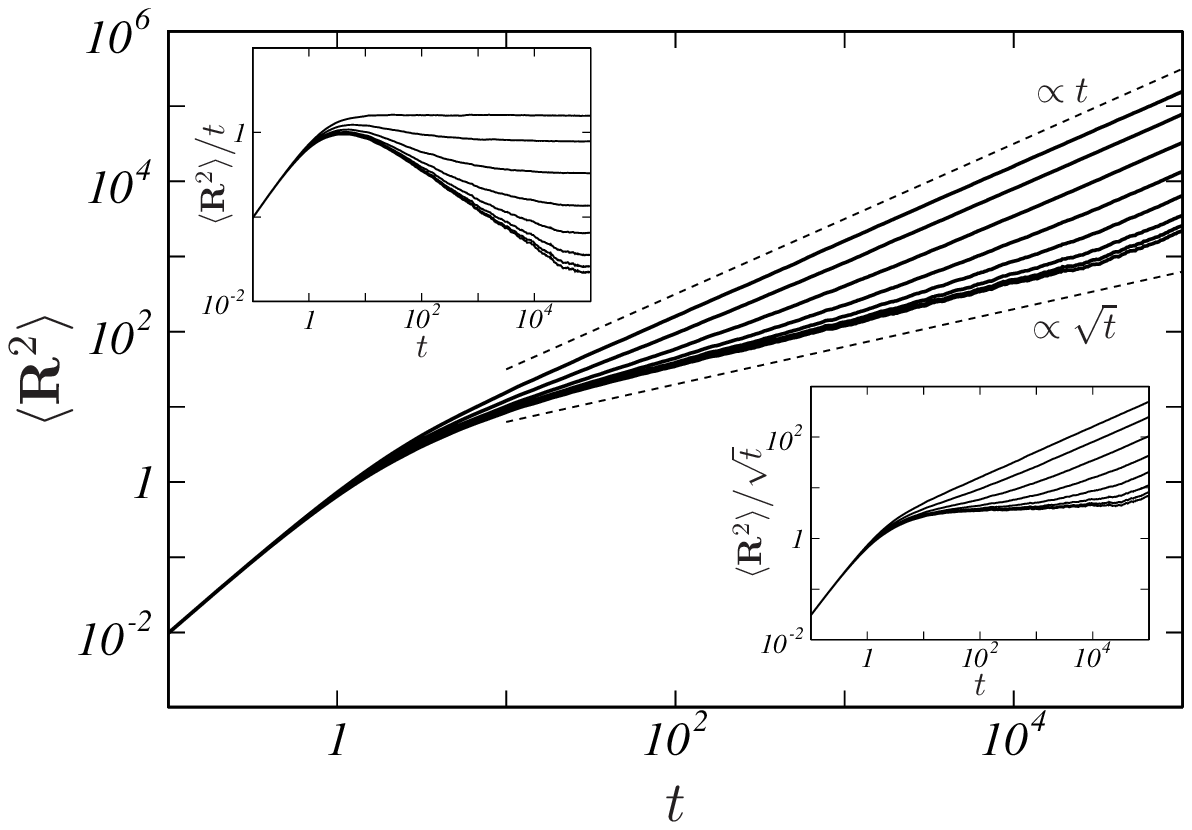}%
  }%
  \caption{\label{Fig:2step}%
    Behavior of the MSD, showing two-step relaxation.
    The time $t$ and the MSD are nondimensionalized 
    with $\sigma^2/D$ and $\sigma^2$, respectively. 
    (a) 
    A three-dimensional system,  
    densely packed with particles 
    subject to Newtonian dynamics with thermostat%
    \protect\cite{Otsuki.PRE86}.
    (b) 
    SFD with overtaking allowed 
    [$\Vmax/\kT = 1, 2, 3, \ldots$, up to $8$; 
    see the text below Eq.~(\protect\ref{Langevin.X})].
    The insets show the MSD compensated by $t$ and $\sqrt{t}$. 
  }
\end{figure}

Suppression of free diffusion in densely packed systems  
  changes the behavior of the MSD significantly.
Typical behavior of the MSD in glassy liquids 
  is depicted in Fig.~\ref{Fig:2step}(a).
After a short period of nearly free diffusion,
  a plateau appears in the MSD, 
  as the particle is trapped in a cage 
  formed by its neighbors.
Later the particle escapes from the cage 
  and the MSD slowly diverges to the infinity.
All this behavior 
  reflects 
  different regimes of relaxation in glassy liquids, 
  referred to as the $\alpha$ and $\beta$ relaxations%
  \cite{Goetze.TTSP24,Kob.PRE51,Kob.PRE52,Reichman.JStat2005}.
The slowest process is the $\alpha$ relaxation, 
  considered to be the structural relaxation 
  that enables the longtime growth of the MSD to infinity. 
Note that this is not a finite-size effect, 
    as we are considering the limit of $L\to+\infty$.
The second slowest process,  
  corresponding to the period 
  in which the MSD crosses the plateau, 
  is termed as the $\beta$ relaxation. 

The slowness of the $\alpha$ and $\beta$ relaxations 
  is expected to have its origin 
  in cooperative motions of the particles,
  correlated both in space and time.
Numerical simulations have revealed  
  spatiotemporally non\-uniform relaxation 
  referred to as dynamical heterogeneity%
  \cite{Yamamoto.PRE58,Berthier.Book2011}, 
  as well as 
  vortical motions\cite{Doliwa.PRE61,Brito.JCP131,Sota.JKPS54}
  and string motions\cite{Donati.PRL80}.
In order to capture such cooperative relaxation processes, 
  various space-time correlations have been devised. 
A popular idea 
  is to define four-point space-time correlations   
  by analogy with spin glasses%
  \cite{Berthier.RMP83,Dasgupta.EPL15,Glotzer.JCP112}, 
  while there are different ideas 
  that involve particle tracking, 
  such as the bond breaking correlation%
  \cite{Yamamoto.PRE58,Shiba.PRE86,Kawasaki.PRE87}.
Combining the idea 
  of continuum-based particle tracking\cite{Ooshida.JPSJ80} 
  with that of four-point space-time correlations\cite{Ooshida.PRE88}, 
  we find  
  that \emph{displacement correlation}%
  \cite{Doliwa.PRE61,Majumdar.PhysicaA177}
  can be quite useful.


The present article 
  aims to give some insights 
  into space-time correlations 
  characterizing cooperativity in glassy liquids, 
  by studying a simpler model of caged dynamics.
By ``a simpler model'' we mean 
  the one-dimensional (1D) version 
  of Eq.~(\ref{Langevin.rvect}), 
  whose behavior is known 
  as \emph{single-file diffusion} (SFD).
An example of SFD with large but finite interaction potential 
  is shown in Fig.~\ref{Fig:2step}(b).
Here the MSD exhibits two-step behavior,
  similar to the one shown in Fig.~\ref{Fig:2step}(a),
  though there is a difference 
  that the ``$\beta$ relaxation'' in standard SFD 
  takes the form of a sub\-diffusive regime 
  with $\Av{R^2} \propto \sqrt{t}$ 
  and not a plateau\footnote{%
      It is known 
      that MSD in some extensions of SFD
      can be slower than $t^{1/2}$; see Subsec.~2.1.%
  }.
For the limiting case 
  of infinite interaction potential, 
  the $\alpha$ relaxation disappears 
  so that the 1D cages become eternal. 
If $L$ is finite, 
  eventually the correlated motion dominates the whole system 
  and the MSD grows again in proportion to $t$, 
  though much more slowly than in the free diffusion%
  \cite{Rallison.JFM186,van-Beijeren.PRB28,%
        Nelissen.EPL80,Tkachenko.PRE82}.  

The article 
  is organized as follows.
Basics of SFD are reviewed 
  in Sec.~\ref{sec:SFD.basic},
  with emphasis on cooperativity  
  rather than anomalous diffusion.
After a survey 
  of certain approaches to SDF 
  in Sec.~\ref{sec:SFD.review},
  we introduce 
  in Sec.~\ref{sec:AP} 
  what we call the Alexander--Pincus (AP) formula%
  \cite{Ooshida.JPSJ80,Ooshida.PRE88,Alexander.PRB18}, 
  to calculate the displacement correlation
  as well as the MSD as its special case.
Improvements 
  over the original version of the formula\cite{Alexander.PRB18}
  are discussed 
  and the formula is extended to the two-dimensional (2D) case, 
  which is the main result of the present work.
Applying the AP formula 
  to the nonlinear dynamics of 1D colloidal liquids,
  we improve the calculation of the MSD 
  in Sec.~\ref{sec:colloid.1D}.
Preliminary results in 2D colloidal liquids  
  are discussed in Sec.~\ref{sec:colloid.2D}.

\section{Basic Properties of Single-File Diffusion}
\label{sec:SFD.basic}

\subsection{Sub\-diffusive behavior}
\label{subsec:SFD.subdif}

SFD is an old problem\cite{Hodgkin.JPhysiol128,Harris.JAP2}
  but remains an active topic. 
In particular, 
  there are a number of works on SFD 
  motivated by interest in glassy dynamics%
  \cite{Ooshida.JPSJ80,Ooshida.PRE88,%
  Rallison.JFM186,Lefevre.PRE72,Abel.PNAS106,Frusawa.PhysLettA378}.  
SFD is also related 
  to a study on polymeric entanglement\cite{Miyazaki.JCP117}
  and the idea of a figure-eight model 
  for kinetic arrest\cite{Pal.PRE78}.

Here we consider the Brownian SFD, 
  governed by the Langevin equation,
\begin{equation}
  m \ddot{X}_i
  = -\mu \dot{X}_i 
  - \dd{}{X_i} \sum_{j<k} V(X_k - X_j)%
  + \mu f_i(t)
  \label{Langevin.X},
\end{equation}
  which is the 1D version 
  of Eq.~(\ref{Langevin.rvect}), 
  with $X_i = X_i(t)$ 
  denoting the position of the $i$-th particle.
The inter\-particle potential $V(r)$
  defines the particle diameter $\sigma$, 
  and $\Vmax = \max{V(r)}$ can be either finite or infinite. 

Denoting the displacement of the $j$-th particle 
  with $R_j(t) = X_j(t) - X_j(0)$, 
  let us review the behavior of the MSD, $\Av{R^2}$, 
  in the over\-damped regime ($t\gg{m/\mu}$).
Setting $V(r) = 0$ results in free diffusion 
  subject to Eq.~(\ref{MSD.free}),
  while the case with $\Vmax = +\infty$, 
  in which overtaking is forbidden,  
  is sub\-diffusive. 
The long-time behavior of the MSD for $\Vmax = +\infty$
  is described by the Hahn--K{\"a}rger--Kollmann (HKK) law%
  \cite{Hahn.JPhA28,Kollmann.PRL90}, 
\begin{equation}    
  \Av{R^2} 
  \simeq \frac{2S}{\rho_0} \sqrt{\frac{{\Dc}t}{\pi}} 
  \label{HKK}, 
\end{equation}
  where $S = S(k\to0)$ 
  is the long-wave limiting value 
  of the static structure factor
\begin{equation}
   S(k) = \frac1N\sum_{i,j}
   \Av{ \exp\left[ \I{k}\left(X_j-X_i\right)\right] }
   \quad (k\ne0)
   \label{S=}, 
\end{equation}  
  and $\Dc = \Dc(k\to0)$ 
  is that of the collective diffusion coefficient, $\Dc(k)$, 
  given by $\Dc = D/S$  
  in the absence of hydrodynamic interaction%
  \cite{Naegele.PhysRep272,Dhont.Book1996}.
Large but finite values of $\Vmax$, 
  as well as finite channel width 
  in quasi-1D systems,  
  allow overtaking%
  \cite{Lucena.PRE85,Siems.SR2,Wanasundara.JCP140,Ooshida@Rhodes1407+}, 
  which introduces the ``$\alpha$ relaxation'' 
  resulting in the two-step behavior in Fig.~\ref{Fig:2step}.
Details of the transitional behavior in quasi-1D systems  
  depend on the softness of the confinement%
  \cite{Lucena.PRE85,Wanasundara.JCP140}. 
The effect of the confinement has been studied 
  through the cross-sectional density profile%
  \cite{Lucena.PRE85}   
  and also through calculation of the free energy barrier 
  from the configurational integral%
  \cite{Wanasundara.JCP140,Ooshida@Rhodes1407+}.

It may be worth noting 
  that there are several extensions of SFD 
  to which the HKK law (\ref{HKK}) does not apply.
The HKK law is premised 
  on statistical uniformity of the system, 
  absence of long-range interactions, 
  and validity of description 
  by the Langevin equation (\ref{Langevin.X})
  with constant $\mu$ and uncorrelated random forcing. 
Without these premises,
  the HKK law may well be modified.
Flomenbom and Taloni\cite{Flomenbom.EPL83}
  studied the case 
  in which the initial distribution of the particles 
  is inhomogeneous,
  in such a way 
  that one tagged particle is located at $X_0 = 0$
  and other particles are arranged around it 
  according to the relation  
  $X_{\pm{j}} \propto \pm j^{1/(1-a)}$
  with $0 \le a \le 1$, 
  so that the density decreases 
  as the distance from the tagged particle increases.
In this case, 
  $\Av{[R_0(t)]^2}$ grows in proportion to $t^{(1+a)/2}$, 
  which interpolates 
  between the standard HKK law (\ref{HKK})
  and Eq.~(\ref{MSD.free}) 
  for free Brownian particles\cite{Flomenbom.EPL83}.
The analysis was extended 
  to the cases 
  of a heterogeneous single file
  with distributed diffusion coefficient\cite{Flomenbom.PRE82}
  and ``anomalous'' single files 
  whose elementary process 
  involves power-law distribution of waiting time%
  \cite{Flomenbom.EPL83,Flomenbom.PRE82,Flomenbom.PLA374}; 
  the MSD in these cases
  can be slower than $t^{1/2}$, 
  and even becomes as slow as $(\log{t})^2$ 
  in extreme cases\cite{Flomenbom.EPL94}.
Some other mechanisms,  
  such as blocking at a junction\cite{Pal.PRE78}
  or spatially correlated noise\cite{Tkachenko.PRE82}, 
  can make the diffusion so slow 
  that a plateau may appear 
  in the logarithmic plot of the MSD.
It is also suspected 
  that the HKK law (\ref{HKK}) may break down
  in regard to longtime behavior of extremely dense single files%
  \cite{Nelissen.EPL80,Frusawa.PhysLettA378}.  
In the following discussion of SFD, however, 
  we will mainly focus on the standard case 
  in which the HKK law is valid.  

For later convenience, 
  we introduce the length scale $\ell_0$,  
  expressing the mean distance between neighboring particles, 
  by $\rho_0 = N/L^\nd = 1/\ell_0^\nd$.
With $\ell_0$ introduced, 
  we also define 
  $D_* \eqdef D/\ell_0^2$ and $\Dc_* \eqdef \Dc/\ell_0^2$, 
  so that $1/\Dc_*$ gives a time\-scale 
  at which $\sqrt{{\Dc}t}$ becomes comparable to $\ell_0$.  
For 1D systems of rigid spheres, 
  we have $\ell_0 = 1/\rho_0 = L/N$
  and $\Dc_*(k) \simeq \rho_0^2 D [1 + (2\rho_0/k)\sin{{\sigma}k}]$.

\subsection{Co\-operativity in single file}
\label{subsec:SFD.co-op}

While many studies on SFD now focus on the anomalous diffusion, 
  it should be emphasized 
  that another aspect of SFD was noticed 
  as early as in 1955,  
  when Hodgkin and Keynes\cite{Hodgkin.JPhysiol128} 
  reported their experiments on potassium ion permeability 
  across the membranes of giant axons 
  from \textit{Sepia officinalis}.
They measured the $\mathrm{K^+}$ fluxes 
  through the membrane  
  in opposite directions,  
  $Q_{\mathrm{{out}\to{in}}}$ and $Q_{\mathrm{{in}\to{out}}}$,
  and compared their ratio 
  with Ussing's equation\cite{Ussing.ActaPhysiolScand19}
  (reminiscent of the fluctuation theorem\cite{Hsieh.BiophysChem139}):
\begin{equation}
  \frac{Q_{\mathrm{{out}\to{in}}}}{Q_{\mathrm{{in}\to{out}}}}
  = \exp\left( \frac{n\Delta\Phi}{\kT} \right)
  \label{Ussing},
\end{equation}
  where 
  $
  \Delta\Phi 
  = q(E_{\text{out}} - E_{\text{in}})
  - \kT\log ({\rho_{\text{in}}}/{\rho_{\text{out}}})
  $
  is the electrochemical potential difference,
  with $q$ denoting the charge of $\mathrm{K^+}$; 
  the electric potential and the ion density 
  on the side $a$ 
  are denoted with $E_a$ and $\rho_a$, respectively. 
Independent diffusion of ions 
  should give $n=1$,  
  while the experimental value was $n = 2.5 > 1$, 
  indicating some kind of cooperativity.
``In a stroke of genius'' (if we borrow 
  the words by Hille\cite{Hille.NM5}), 
  forty years before crystallographic determination 
  of ion channel structure, 
  they said that the experimental result can be explained 
  by assuming that $\mathrm{K^+}$ ions tend to move 
  in narrow channels.

\begin{figure}
  \centering
  \raisebox{0.18\linewidth}{(a)}%
  \IncFig[clip,height=0.20\linewidth]{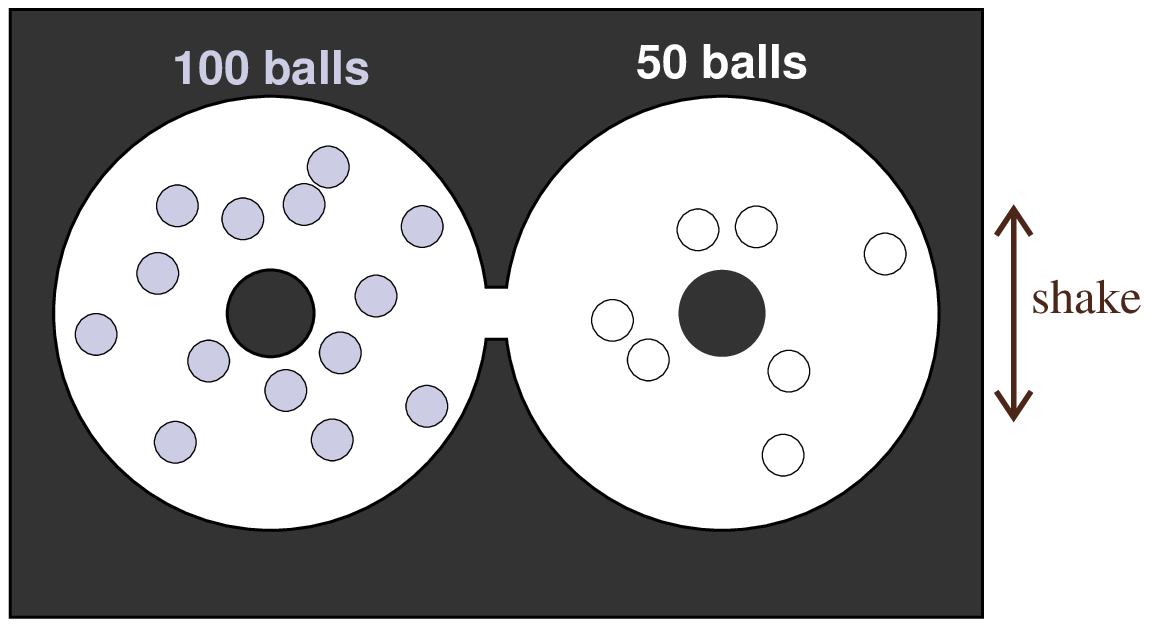}\ 
  \raisebox{0.18\linewidth}{(b)}%
  \IncFig[clip,height=0.20\linewidth]{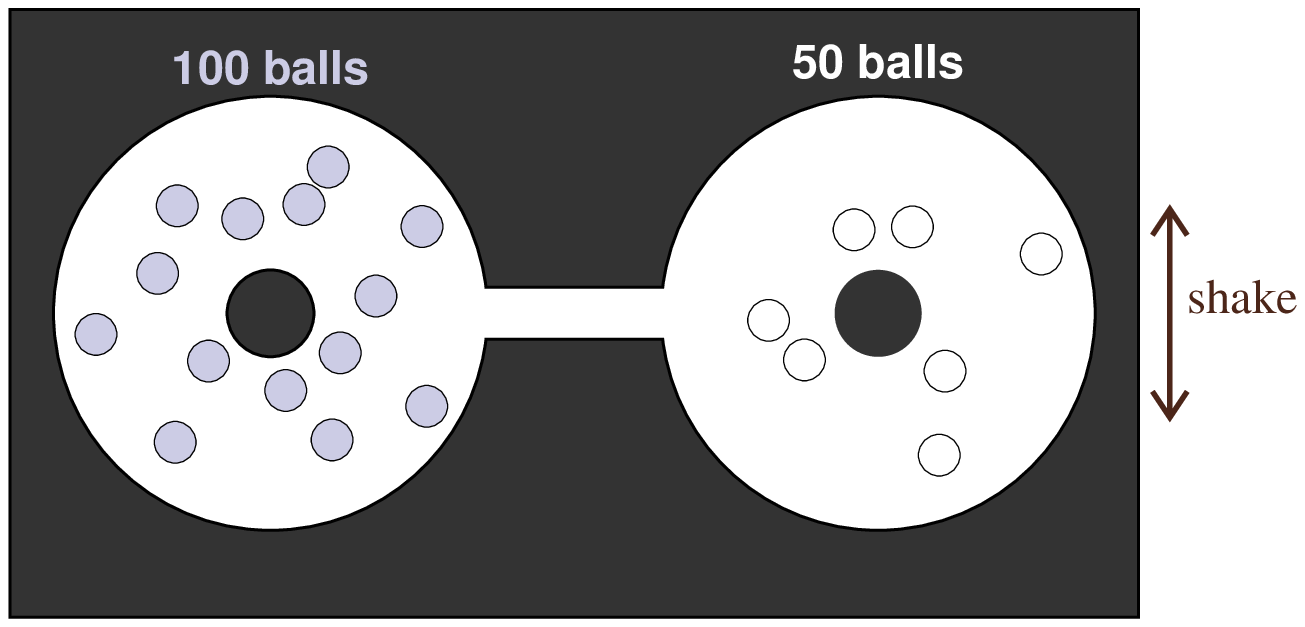}
  \caption{\label{Fig:Hodgkin}%
    Experimental setup of models with granular particles 
    (iron balls with $\sigma = 3\,\mathrm{mm}$)
    to simulate diffusion of ions through a channel;
    adopted from Fig.~8 of Ref.~\protect\refcite{Hodgkin.JPhysiol128}
    with modification.
    The gap length is minimal in (a), 
    while the gap is elongated in (b) to form a long pore.
  }
\end{figure}
  
To demonstrate this idea,  
  Hodgkin and Keynes\cite{Hodgkin.JPhysiol128} 
  performed a mechanical experiment 
  with granular models illustrated in Fig.~\ref{Fig:Hodgkin}.
Causing random motion of particles 
  by shaking the system, 
  they observed fluxes between the chambers 1 and 2 
  and compared their ratio 
  with Eq.~(\ref{Ussing}),  
  which reduces to 
  ${Q_{1\to2}}/{Q_{2\to1}} = (\rho_1/\rho_2)^n$ 
  for electrically neutral particles.
The experiment with a short gap  
  in Fig.~\ref{Fig:Hodgkin}(a)
  gave the ratio 
  close to $\rho_1/\rho_2 = 100/50 = 2$, 
  implying nearly independent diffusion ($n\approx 1$).
Contrastively,  
  for a longer gap in Fig.~\ref{Fig:Hodgkin}(b), 
  $Q_{1\to2}$ was 18 times greater than $Q_{2\to1}$, 
  indicating that about four particles were cooperating
  ($n = \log_2{18} \simeq 4.2$).

In regard to ion transport across membranes, 
  researchers have also investigated possibilities 
  other than the long pore, 
  assuming cooperativity 
  on the side of the transporting proteins 
  and not among the transported ions, 
  which, however, 
  turned out to be unsuccessful\cite{Hill.PNAS68.II}.
The ``knock-on'' mechanism among the $\mathrm{K^+}$ ions 
  is already cooperative, 
  and leads to the high sensitivity of the flux ratio
  to $\Delta{\Phi}$.
In this case, the system size is finite, 
  and the correlation length can span the entire system.


\begin{figure}
  \centering
  \raisebox{0.30\linewidth}{(a)}\!%
  \raisebox{2.5em}{%
  \IncFig[width=0.35\linewidth]{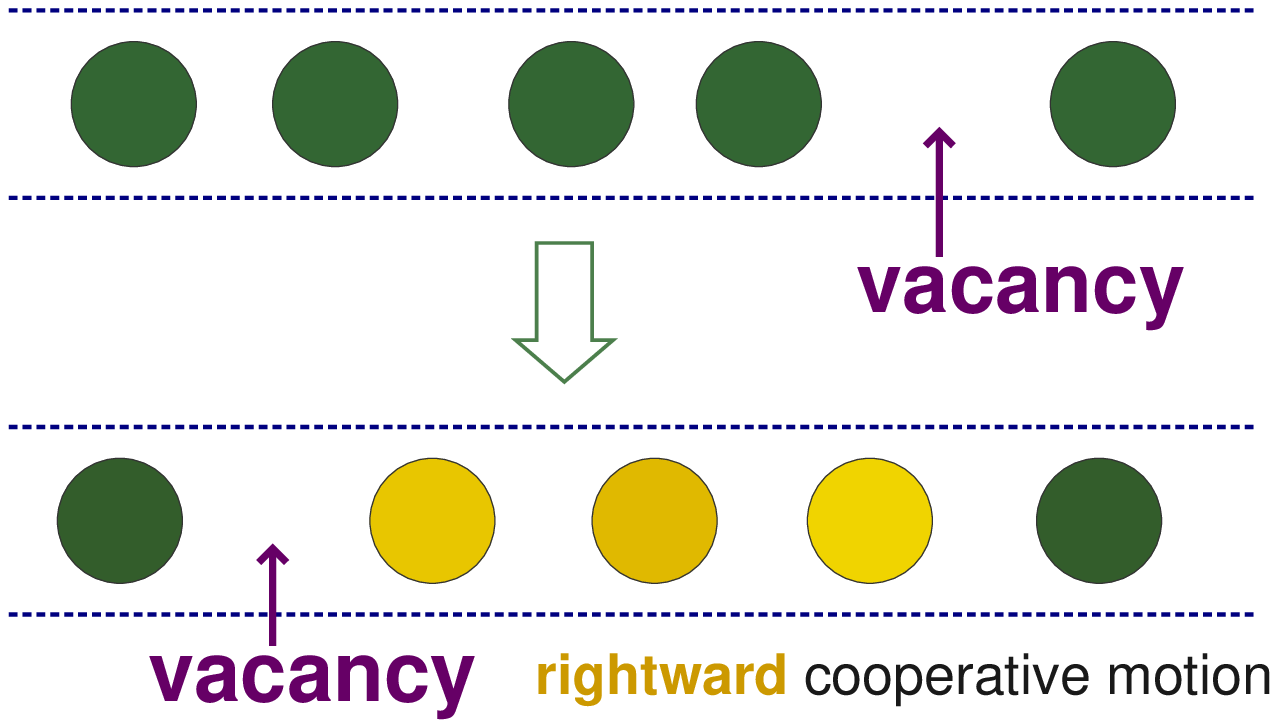}}%
  \qquad
  \raisebox{0.30\linewidth}{(b)}\!%
  \IncFig[width=0.45\linewidth]{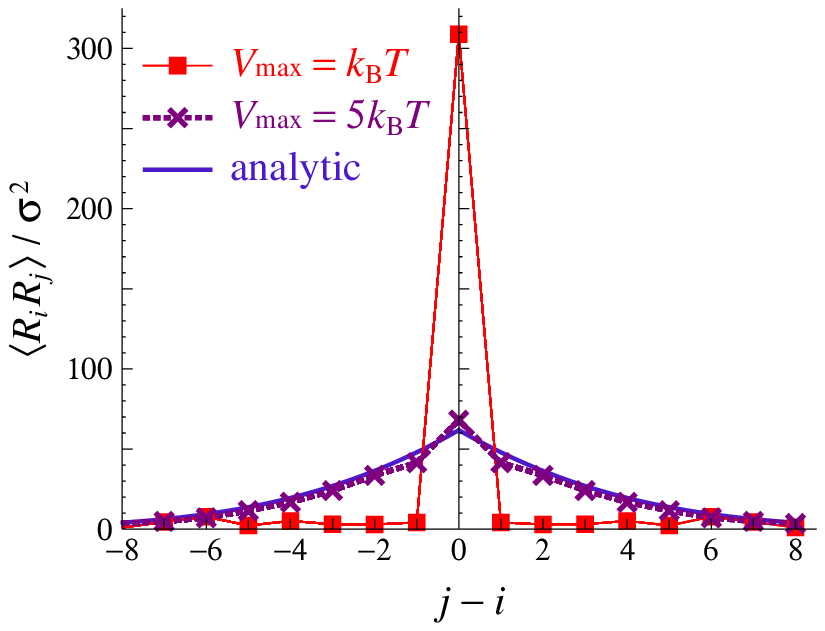}%
  \caption{\label{Fig:DC.1D}%
     Cooperative motion in 1D systems. 
     (a) 
     Schematic description of cooperative motion 
     due to a diffusing vacancy. 
     (b) 
     Numerically calculated 
     two-particle displacement correlation $\Av{{R_i}{R_j}}$ 
     in SFD 
     for different values of $\Vmax$, 
     with $N = 10^4$, $\rho_0 = N/L = 0.2\,\sigma^{-1}$,
      and the time interval $t = 200\,\sigma^2/D$;
      cited from Ref.~\protect\refcite{Ooshida@Rhodes1407+}.       
      The analytical solution is given by Eq.~(\ref{R*R}). 
  }
\end{figure}


In situations with $L\to+\infty$, 
  including the case 
  in which the HKK law (\ref{HKK}) holds, 
  the cooperativity exhibits itself 
  through slow diffusion. 
Cooperativity 
  is usually detected 
  by comparing two configurations at different times, 
  as is illustrated in Fig.~\ref{Fig:DC.1D}(a).
Here, 
  the leftward diffusion of the vacancy 
  induces rightward cooperative displacement of particles.
This cooperative motion  
  can be 
  quantified by calculating 
  displacement correlation%
  \cite{Ooshida.PRE88,Majumdar.PhysicaA177,Ooshida@Rhodes1407+}, 
  given by $\Av{{R_i}{R_j}}$
  in the 1D case. 

A numerical example
  of displacement correlation\cite{Ooshida@Rhodes1407+}  
  is shown in Fig.~\ref{Fig:DC.1D}(b).
The cases with $\Vmax = \kT$ and $\Vmax = 5\kT$ 
  are compared.  
After sufficiently long equilibration, 
  the particles were renumbered consecutively at $t=0$, 
  and the values of $\Av{{R_i}{R_j}}$ 
  were calculated for $t = 200\,\sigma^2/D$.
In the case of relatively low barrier ($\Vmax = \kT$),  
  the displacements of the neighboring particles 
  are nearly uncorrelated, 
  indicating that the particles are diffusing almost freely.
Contrastively, 
  in the case of higher barrier ($\Vmax = 5\kT$),
  positive correlation is observed 
  within some finite distance.
This correlation 
  demonstrates dynamical structure 
  behind the slow diffusion.
The dynamical correlation length 
  is diffusive and given by $\lambda(t) = 2 \sqrt{{\Dc}t}$.

\section{Theories of SFD: extensible to higher dimensions?}
\label{sec:SFD.review}

There are 
  various theoretical approaches to SFD.
Some techniques,  
  such as 
  the Jepsen line\cite{Barkai.PRL102,Barkai.PRE81,Leibovich.PRE88}, 
  highly depend on the 1D geometry,
  while other techniques may be extensible 
  to three-dimensional (3D) colloidal liquids.
We look for the latter kind of approaches,  
  requiring in addition 
  that the theory should incorporate cooperativity.

For the sake of simplicity,  
  in the present section, 
  sometimes we 
  ignore the difference between $D$ and $\Dc=D/S$. 
This is strictly valid 
  only in the limit of point particles 
  ($\sigma \ll \ell_0 = L/N$),
  but it is often possible  
  to take the effect of finite $\sigma$ into account 
  by reviving $S$ in an obvious way.

\subsection{Phenomenological treatments of cooperative diffusion}
\label{subsec:phenomen}

Let us start 
  with reviewing the idea of Rallison\cite{Rallison.JFM186}
  that relates the slow diffusion to cooperativity.
If $n$ Brownian particles 
  are strongly interacting 
  and moving together, 
  the diffusion coefficient
  for their center of mass is $D/n$,
  in the sense that 
\begin{equation} 
  \frac{\D}{\D{t}}\Av{R^2} = \frac{2D}{n} 
  \label{2D/n}.
\end{equation}
To apply Eq.~(\ref{2D/n})
  to the collective dynamics in SFD, 
  Rallison\cite{Rallison.JFM186} 
  proposed to replace $n$ in the denominator 
  with $\mathcal{N}(\ld) = 1 + \rho_0\ld$,
  which is the number of particles 
  within the dynamical correlation length $\ld = \ld(t)$,
  so that the MSD is given by 
\begin{equation}  
  \Av{R^2}
  = \int_0
  \frac{2D\D{t}}{\mathcal{N}(\lambda)},
  \quad
  \lambda = \lambda_\text{Ral}^\text{1D}(t) 
  = \sqrt{{4\pi}{\Dc}t}
  \label{Rallison.Eq5}.
\end{equation}
Upon integration, 
  Eq.~(\ref{Rallison.Eq5}) yields
\begin{equation}
  \Av{R^2} 
  = \frac{2S}{\rho_0} \sqrt{\frac{{\Dc}t}{\pi}} 
  - \frac{S}{\pi\rho_0^2} 
    \log\,\left(1 + \rho_0\sqrt{{4\pi}{\Dc}t}\,\right) 
  \label{MSD.Rallison};
\end{equation}  
  it gives free diffusion for small $t$,
  and  
  sub\-diffusion for large $t$ 
  with a logarithmic correction term.

It is interesting to test Rallison's idea 
  in the case of a single file
  with non-uniform initial condition\cite{Flomenbom.EPL83}, 
  given by $X_0(0) = 0$ and 
  $X_{\pm{j}}(0) \propto \pm j^{1/(1-a)}$.
Since 
  $\mathcal{N}(\ld) \propto \ld^{1-a} \propto t^{(1-a)/2}$
  in this case,
  Eq.~(\ref{Rallison.Eq5}) readily yields 
  $\Av{R^2} \propto t^{(1+a)/2}$.
This result 
  was derived by Flomenbom and Taloni\cite{Flomenbom.EPL83}
  with an analogous but slightly different approach, 
  in which $\mathcal{N}(\abs{R})$ was considered 
  instead of $\mathcal{N}(\ld)$.

Rallison\cite{Rallison.JFM186} 
  applied this idea to the 3D colloidal liquid
  and obtained a critical volume fraction, 
  $\phi_{\text{c}} = 0.64$
  (with $\phi$ related to $\rho_0$ 
  by $\phi = {\frac43}\pi(\frac{\sigma}{2})^3\rho_0$),
  above which the MSD cannot grow to infinity.
This is greater 
  than the experimental value of glass transition point
  ($\phi_{\text{glass}} = 0.56$ 
  according to Pusey and van Megen\cite{Pusey.PRL59}), 
  and rather near to the random close packing density.
The discrepancy 
  casts doubt on quantitative validity of the theory,
  but the idea of relating the slow diffusion 
  to the number of particles in cooperative motion 
  remains quite suggestive.

Another instructive idea 
  is found in the lattice SFD  
  by van Beijeren \textit{et al.}\cite{van-Beijeren.PRB28}.
On the basis   
  of the picture of migrating vacancies, 
  i.e.\ 
  the discrete version of Fig.~\ref{Fig:DC.1D}(a), 
  they calculated velocity autocorrelation 
  of the tagged particle as 
\begin{equation}
  \Av{u(t)u(0)} \simeq - A_u t^{-3/2}
  \label{Beijeren83.VAF}
\end{equation}
  for large $t$, 
  where $u$ denotes the velocity 
  and $A_u$ is a constant 
  that depends 
  on the hopping rate, concentration, and lattice constant.
With the aid 
  of the relation 
\begin{equation}
  \frac{\D^2\Av{R^2}}{\D{t}^2} 
  = 2 \Av{u(t)u(0)}
  \label{MSD//VAF},
\end{equation}
  Eq.~(\ref{Beijeren83.VAF}) 
  readily leads to the asymptotic law, 
  $\Av{R^2} \propto \sqrt{t}$.

The velocity autocorrelation in Eq.~(\ref{Beijeren83.VAF}) 
  has a negative longtime tail: 
  the particles are always pushed back by its neighbors.
The negative tail 
  can be derived 
  also on the basis of continuum description\cite{Taloni.PRE78},  
  as will be reviewed later in Eq.~(\ref{Taloni.Eq16}).

Within the framework of lattice dynamics, 
  the idea of migrating vacancies
  can be applied also to geometries 
  other than the 1D chain, 
  such as double-file diffusion\cite{Kutner.PRB30}
  and 2D regular lattices\cite{van-Beijeren.PRL55}.
The idea of clarifying slow dynamics by moving defects 
  has been incarnated 
  also in free-volume kinetic models 
  of glasses and granular matters%
  \cite{Berthier.RMP83,Sellitto.PRE62}. 
An interesting question 
  is whether this approach can be generalized 
  to systems without lattice,  
  such as the one shown in Fig.~\ref{Fig:DC.1D}(a).
The answer is affirmative in our opinion, 
  as will be explained in the latter half of this article.

The theory of Rallison\cite{Rallison.JFM186}
  and that of van Beijeren \textit{et al.}\cite{van-Beijeren.PRB28}
  were immediately applied to SFD on a ring with finite $L$.
The longtime behavior in this case 
  is given by Eq.~(\ref{2D/n}) with $n=N$.

\subsection{Density correlation in the Fourier space}
\label{subsec:Fourier}

As the dynamical correlation length in Brownian SFD 
  is diffusive, 
  it is natural to seek indication of collective motions 
  in the density field 
  subject to a stochastic diffusion equation.
Here we review several ideas  
  for relating the motion of a tagged particle
  to the density fluctuations of the medium, 
  including the celebrated theory  
  by Alexander and Pincus\cite{Alexander.PRB18}.
  
Let us define 
  the (microscopic) density field by
\begin{equation}
  \rho = \rho(x,t) = \sum_i \rho_i 
  \relax, \quad 
  \rho_i = \rho_i(x,t) = \delta(x-X_i(t)) 
  \label{rho=}
\end{equation}
  where
  $\rho_i$ is the single-body density of the $i$-th particle.
Evidently, 
  $\rho$ satisfies the continuity equation
\begin{equation}
  \dt\rho + \dx{Q} = 0  \label{cont1D},
\end{equation}
  with the flux $Q = {\rho}u$ 
  given by 
\begin{equation}
  Q = Q(x,t) = \sum_i \rho_i(x,t) \dot{X}_i(t)  \label{Q=}.
\end{equation}

Anticipating a linear equation for $\rho(x,t)$  
  as a starting point 
  and assuming 
  that the system is translationally invariant, 
  we introduce the Fourier representation:
\begin{equation}
  \hat\rho(k,t) 
  = {\frac1L} \int e^{{\I}kx} 
  \left[ \rho(x,t) - \rho_0 \right]\D{x}
  \relax, \quad 
  \rho(x,t) 
  = \rho_0 + \sum_k \hat\rho(k,t) e^{-{\I}kx}
  \label{k1+3}, 
\end{equation}
  with $k/\Delta{k} \in \Z$ where $\Delta{k} = 2\pi/L$.

The two-time correlation of $\hat\rho$  
  is referred to 
  as the intermediate scattering function 
  and essentially equivalent 
  to the dynamical structure factor\cite{Hansen.Book2006},
  as  
\begin{equation}
   F(k,t) = \frac1N\sum_{i,j}
   \Av{ e^{-\I{k}\left[X_i(t) - X_j(0) \right]} }
   \propto \Av{{\hat\rho(k,t)}{\hat\rho(-k,0)}}
   \quad
   (k \ne 0)
   \label{F=};
\end{equation}  
  the self part of $F$ gives 
  the Fourier representation of the diffusion propagator,
\begin{equation}
  \Fs(k,t) 
  = L^2 \Av{ \hat\rho_j(k,t) \hat\rho_j(-k,0) } 
  = \Av{ e^{\I{k}{R_j(t)}} }
  \label{Fs=}, 
\end{equation}
  where 
  $\hat\rho_j(k,t) = L^{-1} \exp\left[ \I k X_j(t) \right]$
  is the Fourier component of $\rho_j(x,t)$.
  
Now let us 
  proceed to the dynamics of $\rho$.
If the particles 
  are subject to the Langevin equation (\ref{Langevin.X}) 
  and attention is focused 
  on the timescale longer than $m/\mu$, 
  then $\rho$ is governed 
  by a nonlinear stochastic equation 
  referred to as the Dean--Kawasaki equation%
  \cite{Dean.JPhAMG29,Kawasaki.PhysicaA208,Kawasaki.JStatPhys93}.
Here we discuss 
  the linear dynamics only,
  deferring the nonlinear case 
  until Sec.~\ref{sec:colloid.1D}.
The linear stochastic equation for $\rho$,  
  sometimes referred to as the diffusion-noise equation%
  \cite{Taloni.BRL9,Taloni.PRE78},   
  reads 
\begin{equation}
  \dt\rho(x,t) = D \dx^2 \rho + f_\rho(x,t)
  \label{diffusion-noise}, 
\end{equation}
  with the statistics of the noise term given by 
\begin{align}
  \Av{f_\rho(x,t) f_\rho({x'},{t'})}
  &= 
  2D \dx \partial_{x'} \rho(x,t) \delta(x-x') \delta(t-t') 
  \notag \\   
  &\simeq 
  2\rho_0 D \dx \partial_{x'} \delta(x-x') \delta(t-t') 
  \label{f2.approx}.
\end{align}%

The linearized Dean--Kawasaki equation (\ref{diffusion-noise}) 
  is readily solved in the Fourier representation.  
With this solution, 
  Taloni and Lomholt\cite{Taloni.PRE78} 
  calculated the velocity autocorrelation, 
  $\Av{{u_j(t)}{u_j(0)}}$ with $u_j = \dot{X}_j$,
  which is related to the MSD by Eq.~(\ref{MSD//VAF}).
The single-file condition implies 
  that the velocity of the tagged particle 
  equals that of the medium around it 
  (for the time\-scale longer than $1/D_*$), 
  so that 
\begin{equation}
  u_j(t) = u(X_j(t),t) 
  \simeq 
  u(X_j(0),t) = \frac{Q(X_j(0),t)}{\rho_0}
  \relax.
\end{equation}
The replacement of $X_j(t)$ by $X_j(0)$
  is justified 
  by the relative smallness of the displacement 
  in comparison to the dynamical correlation length. 
Using the continuity equation (\ref{cont1D}) 
  and the space-time Fourier transform 
  of the noise term,  
  they found\cite{Taloni.PRE78} 
\begin{equation}
  \Av{{u(t)}{u(0)}}
  = -\frac{1}{4\rho_0}\sqrt{\frac{D}{{\pi}t^3}}
  \label{Taloni.Eq16},  
\end{equation}
  reproducing the same negative longtime tail 
  as in Eq.~(\ref{Beijeren83.VAF}).
We note 
  that the negative tail in the velocity autocorrelation 
  is a manifestation of the cage effect, 
  found in 2D and
  3D colloidal systems 
  as well\cite{Dhont.Book1996,Hagen.PRL78},
  and differs from the positive longtime tail 
  reported by Alder and Wainwright\cite{Alder.PRA1}.

Taloni and Lomholt\cite{Taloni.PRE78} also succeeded 
  in interpolating 
  the longtime collective dynamics in SFD 
  and the short-time single-particle behavior, 
  describing the velocity of the particle 
  with a generalized Langevin equation
\begin{equation}
  m\frac{\D{u}}{\D{t}} 
  = -\int_0^t K(t-t') u(t') \D{t'} 
  + f_K(t)
  \label{gLangevin}, 
\end{equation}
  where $f_K$ is a time-correlated noise 
  related to the kernel $K$ 
  by the relation 
\begin{equation}
  \Av{{f_K(t)}{f_K(t')}} = {m\kT} K(\abs{t-t'})
  \label{fK}.
\end{equation}
The longtime behavior of $K$ can be determined 
  by Fourier analysis 
  of the collective dynamics in Eq.~(\ref{diffusion-noise}) 
  or its chain-dynamical equivalent%
  \cite{Taloni.BRL9,Taloni.PRE78,Lizana.PRE81}.
Furthermore, 
  the short-time behavior 
  should reduce to that of a free Brownian particle.
By incorporating these two limiting cases into $K$, 
  Taloni and Lomholt\cite{Taloni.PRE78} obtained 
\begin{equation}
  \Av{R^2} 
  \simeq 
  \frac{1}{\rho_0^2}
  \left[
  2\sqrt{\frac{{D_*}t}{\pi}}
  + \frac{e^{4{D_*}t}}{2} \erfc\left(2\sqrt{{D_*}t}\,\right)
  - \frac12
  \right]
  \label{MSD.Taloni}. 
\end{equation}

Contrastively, 
  the concise theory 
  by Alexander and Pincus\cite{Alexander.PRB18}
  avoids directly dealing with the velocity $u$
  and relates the MSD 
  directly to the density fluctuation,  
  by the integral 
\begin{equation}
  \Av{R^2}
  \propto 
  \int \frac{1 - e^{-Dk^2t}}{k^2}\D{k}
  \label{AP.original}.
\end{equation}
We refer to Eq.~(\ref{AP.original}) and its extensions 
  as the Alexander--Pincus (AP) formula.
  
The original derivation of Eq.~(\ref{AP.original})  
  is traced as follows:  
Let $h_j = h_j(t)$ 
  denote the displacement of the $j$-th particle 
  from a configuration 
  in which the particles are uniformly distributed, 
  so that $h_j(t) = X_j(t) - j\ell_0$.
Rewriting $\hat\rho(k,t)$ with $h_j$ 
  as
\begin{equation}
  \hat\rho(k,t) 
  = \sum_j \hat\rho_j(k,t) 
  = {\frac1L}\sum_j 
  e^{ \I{k}\left[ j\ell_0 + h_j(t) \right] }
  \quad
  (k\ne0)
  \label{k1*} 
\end{equation}
  and introducing 
  $\hat{h}(k,t) = {\frac1N} \sum_j e^{{\I}kj\ell_0} h_j(t)$
  so that
  $ h_j(t) = \sum_k \hat{h}(k,t) e^{-{\I}kj\ell_0} $, 
  one expands $\hat\rho(k,t)$ in Eq.~(\ref{k1*}) 
  in power series of $k h_j(t)$, 
  to find 
\begin{equation}
  \hat\rho(k,t)
  = 
  {\frac1L} \sum_j e^{\I{k}j\ell_0} 
  \left[ 1 + {\I}k h_j(t) + O(k^2 h_j^2) \right]  
  \simeq {\I}k\rho_0 \hat{h}(k,t)
  \label{AP.Eq2}.
\end{equation}
By noticing $R_j(t) = h_j(t) - h_j(0)$ 
  and using Eq.~(\ref{AP.Eq2}),   
  the MSD is calculated as  
\begin{equation}
  \Av{R_j^2}
  = 
  2\left\{
  \Av{\smash{\left[h_j(0)\right]^2}} - \Av{h_j(t)h_j(0)}
  \right\} 
  \propto 
  \sum_k \frac{F(k,0) - F(k,t)}{k^2} 
  \label{AP.Eq4}; 
\end{equation}
  then, 
  replacing the summation with an integral 
  and substituting  
  $F(k,t) \propto e^{-Dk^2t}$,
  we arrive at Eq.~(\ref{AP.original}), 
  from which $\Av{R^2} \propto \sqrt{t}$ is readily obtained.

It should be noted 
   that the above derivation 
  includes a step requiring attention.
Since $h$ is not small,  
  the validity of the expansion in Eq.~(\ref{AP.Eq2}) 
  poses a delicate issue. 
However, 
  this difficulty is avoided 
  by using a convected coordinate system, 
  as will be shown in Sec.~\ref{sec:AP}.

\subsection{Mode-coupling theory}
\label{subsec:MCT}

Conceptually, 
  the generalized Langevin equation (\ref{gLangevin}) 
  could be regarded 
  as projection of the $N$-body equation (\ref{Langevin.X}) 
  onto the single-particle motion. 
However, this projection 
  seems to be difficult to perform systematically 
  in the real space 
  (as oppose to the Fourier space)\cite{Taloni.BRL9}.
It is then interesting 
  to notice the case 
  in which 
  the Mori--Zwanzig projection formalism
  has been applied 
  to 3D dynamics of glassy liquids, 
  known by the name of mode-coupling theory (MCT)%
  \cite{Reichman.JStat2005,Gotze.RPP55,Goetze.Book2009}.
Projection of the equation of motion 
  onto the density correlation $F$
  yields 
\begin{equation}
  \left(\dt + \Dc k^2\right)  F(k,t) 
  = -\int_0^t  M(k,t-t') \partial_{t'} F(k,t') \D{t'} 
  \label{MCT.F}, 
\end{equation}
  which is then closed 
  by approximating the memory kernel $M$
  with a quadratic functional of $F$.
In spite of various limitations,
  the MCT equation 
  has succeeded 
  in reproducing the two-step relaxation of $F(k,t)$
  and the growth of the relaxation time,
  as the mean density $\rho_0$ increases%
  \cite{Kob.PRE51,Kob.PRE52,Reichman.JStat2005}.

Diffusion of a tagged particle 
  is treated by the MCT equation for $\Fs$, 
  in the form
\begin{equation}
  \left(\dt + D k^2\right)  \Fs(k,t) 
  = -\int_0^t \Ms(k,t-t') \partial_{t'} \Fs(k,t') \D{t'} 
  \label{MCT.Fs},
\end{equation}
  with the memory kernel $\Ms$ 
  expressed as a bilinear functional of $F$ and $\Fs$.
In description of SFD with Eq.~(\ref{MCT.Fs}), 
  however, 
  it is difficult to reproduce the $\sqrt{t}$ behavior.
As long as the standard memory kernel is used, 
  MCT cannot predict anomalous diffusion\cite{Abel.PNAS106}.
Mathematical difficulty is located 
  in constructing $\Ms$ with proper longtime behavior, 
  which is actually more difficult 
  than constructing $K$ in Eq.~(\ref{gLangevin}), 
  because $K$ can be given before the unknown $u$ is determined,  
  while $\Ms$ must be a functional of the unknown $F$ and $\Fs$.
In modified MCT approaches 
  by Fedders\cite{Fedders.PRB17} 
  and Abel \textit{et al.}\cite{Abel.PNAS106},
  $\Ms$ is improved 
  by taking the prohibition of positional exchanges 
  directly into account;
  as a result, 
  the correct $t^{1/2}$-law is recovered.

Another possible approach 
  consists of retaining Eq.~(\ref{MCT.F}) for $F$ alone,
  while Eq.~(\ref{MCT.Fs}) for $\Fs$ is replaced 
  with another equation for tracers,  
  in which the four-point correlation 
  is directly taken into account%
  \cite{Miyazaki.JCP117,Kollmann.PRL90}.
We will see 
  that an improved version of the AP formula 
  is suitable for this purpose\cite{Ooshida.PRE88}.

\subsection{Elastic chains}
\label{subsec:elastic}

The longtime dynamics of SFD 
  are known to be equivalent 
  to those of fluctuating elastic chains.
The linear dynamics are then described 
  by replacing $V(X_k-X_j)$ in Eq.~(\ref{Langevin.X})
  with the harmonic effective potential, 
  so that the equation of motion reads
\begin{equation}
  m \ddot{X}_i + \mu\dot{X}_i 
  = \kappa\left( X_{i+1} - 2 X_i + X_{i-1} \right) 
  + \mu f_i(t) 
  \label{chain}
\end{equation}
  with $\kappa$ denoting the effective spring constant%
  \cite{Lizana.PRE81,Centres.PRE81}.
Taking the continuous limit 
  and considering the over\-damped case ($m\to0$)
  for simplicity,  
  we rewrite Eq.~(\ref{chain}) 
  in terms of $x(\xi,t)$,
  with $\xi$ here denoting a continuum analogue of $i$:
\begin{equation}
  \dt{h(\xi,t)} 
  = D' \dXi^2{h(\xi,t)} + f_h(\xi,t) 
  \label{EW}, 
\end{equation}
  where $D' = \kappa/\mu$ and 
  $
  \Av{{f_h(\xi,t)}{f_h(\xi',t')}} 
  = 2D \delta(\xi-\xi') \delta(t-t')$.
Note that $h$ can be taken 
  equal to the position $x$ itself, as $h=x(\xi,t)$, 
  or the displacement from the mean position, 
  as $h = x(\xi,t) - \ell_0\xi$.
In any case, 
  the displacement is given 
  by $R(\xi,t) = h(\xi,t) - h(\xi,0)$, 
  and the MSD is given by the essentially same integral 
  as Eq.~(\ref{AP.original}). 
The Langevin equation for elastic chains 
  is known as the Rouse model 
  in the context of polymer dynamics\cite{Doi.Book1986}.
The dynamics of a curve $h=h(\xi,t)$ on the $(\xi,h)$ plane 
  can also be interpreted 
  as kinetic roughening of a fluctuating surface%
  \cite{Majumdar.PhysicaA177,Majumdar.PRB44},
  with Eq.~(\ref{EW}) 
  referred to as the Edwards--Wilkinson equation%
  \cite{Edwards.PRSLA381,Krug.AdvPhys46}.

While this reduction to elastic chain 
  seems obvious 
  in the case of soft-core interaction, 
  the important point is 
  that also the hardcore interaction
  can be described by a smooth effective potential 
  between the particles,
  as a result of coarse-graining in time.
Recently, 
  this point has been discussed
  both in the 1D setup\cite{Lizana.PRE81}
  and in the context of glassy liquids 
  in higher dimensions\cite{Brito.JCP131,Brito.EPL76}\relax,
  seemingly by independent groups.
The effective potential, 
  calculated from the partition function 
  that gives the free energy,  
  is of entropic origin:  
  the inter\-particle force 
  is essentially 
  the osmotic pressure of the ideal solution,
  i.e.\
  the pressure of the ideal gas. 


The discrete equation (\ref{chain})
  and its higher-dimensional analogues 
  are amenable to normal mode analysis%
  \cite{Coste.BRL9,Brito.JCP131,Sota.JKPS54}, 
  which is often performed around some metastable configuration.
Modes with low frequencies 
  are found to be important.
In the case of colloidal glass with $\nd\ge2$, 
  the presence of the low-frequency modes  
  is related to the character 
  of the interaction described by the ``bonds'', 
  which can directly sustain compression 
  but not necessarily resist shear deformation.
In the continuum description,  
  the normal modes are plane waves.
In correspondence to the low-frequency modes in discrete systems, 
  the continuum description of colloidal liquids
  must respect 
  the distinction between the longitudinal and transverse modes.

The situation can be simpler 
  in systems that are elastic by nature, 
  such as an assembly of densely packed soft\-core particles 
  or a system of colloidal particles 
  trapped in a polymeric network. 
In such models, 
  the presence of finite rigidity could be taken for granted, 
  so that simplified description 
  based on the scalar integral
\begin{equation}
  \Av{R^2}
  \propto 
  \int \frac{1 - e^{-Dk^2t}}{k^2}\,\D^\nd{\mb{k}}
  \qquad 
  (\text{without tensorial treatment})
  \label{EW.nD}
\end{equation}
  might be justifiable. 
Once the description by the linear elastic model is justified, 
  it allows analytical calculation of displacement correlation
  and even some four-point correlations 
  such as dynamical susceptibility\cite{Toninelli.PRE71}.
Details 
  will be discussed elsewhere.

   
The effect of dimensionality on the MSD in elastic models    
  may deserve some comments.
The 1D AP formula
  is free from ultraviolet divergence, 
  while $t$ must be finite 
  to avoid infrared divergence of the MSD. 
In the 2D and 3D cases,  
  a ultraviolet cutoff must be introduced. 
Denoting 
  the cutoff wavenumber with $k_0 = \tilde{k}_0/\ell_0$,
  we calculate the MSD in the 2D case 
  from Eq.~(\ref{EW.nD})
  as 
\begin{equation}
  \Av{\mb{R}^2} 
  \propto
  \log\left( 1 + \tilde{k}_0^2{\Dc_*}t \right)
  \label{MSD.2D},  
\end{equation}
  and for $\nd=3$ we find 
\begin{equation}
  \Av{\mb{R}^2} 
  \propto 
  1 - \frac{1}{\sqrt{1 + \tilde{k}_0^2{\Dc_*}t}}
  \label{MSD.3D}.
\end{equation}
The large $t$ behavior of Eq.~(\ref{MSD.2D}) 
  differs from that of Eq.~(\ref{MSD.3D}):
  the MSD for $\nd=3$
  converges for $t\to+\infty$, 
  while that for $\nd=2$
  diverges logarithmically, 
  until the effect of the finite system size 
  comes into play at $t\sim L^2/\Dc$.
Thus the 2D elastic model 
  suffers both infrared and ultraviolet divergences.
It can be even argued  
  that the 2D harmonic solid 
  is not a solid in the usual sense\cite{Jancovici.PRL19}, 
  at least if one accepts the argument 
  that single-file systems are denied to be solid 
  because any particle in the system 
  can get arbitrarily far away from its initial position%
  \cite{Diamant.arXiv1406}.

\section{Alexander--Pincus Formula in Convected Coordinate System}
\label{sec:AP}

We have reviewed 
  several theories of SFD  
  that may be extensible to higher dimensions, 
  including the AP formula (\ref{AP.original}), 
  which relates the MSD to $F(k,t) \propto e^{-D{k^2}t}$.
Since the original derivation of the formula 
  is of approximate nature, 
  it seems difficult to improve the result
  by considering a nonlinear correction to $F$.

Here we redevelop the AP formula, 
  making it exact 
  by changing the variables appropriately%
  \cite{Ooshida.JPSJ80,Ooshida.PRE88}. 
The key to this approach 
  is the adoption of Lagrangian correlations,  
  motivated by theories of the Navier--Stokes turbulence%
  \cite{Kraichnan.PhF8,Kaneda.JFM107,Kida.JFM345}.
While the formula itself remains linear 
  with regard to the correlation in the new variables,  
  it can be combined
  with a nonlinear equation of the correlation, 
  so that the result can be improved 
  as we will see later in Subsec.~\ref{subsec:colloid.1D.NL}.
The formula is also extended 
  so as to give the displacement correlation
  in $\nd$-dimensional systems.

\subsection{Label variable and the 1D AP formula}

Let us start 
  with reviewing a classic idea of the Lagrangian description 
  in fluid mechanics. 
The idea is to introduce 
  a continuous and invertible mapping
\begin{equation}  
  (\BoldXi,t) = (\xi,\eta,\zeta,t) 
  \mapsto \mb{r} 
  = \ThreeVect{x(\BoldXi,t)}{y(\BoldXi,t)}{z(\BoldXi,t)}
  \label{L.map}, 
\end{equation}  
  such that its inverse, $\BoldXi = \BoldXi(\mb{r},t)$, 
  satisfies
  the convective equation
\begin{equation}
  \left( \dt + \mb{u}\cdot\nabla \right) \BoldXi(\mb{r},t) 
  = \mb{0}    \label{convect}.
\end{equation}
This mapping makes it possible 
  to specify a world line 
  as ${\BoldXi(\mb{r},t)} = \text{const.}$, 
  and therefore we refer to $\BoldXi$ 
  as the \emph{label variable}.
In other words, 
  $\BoldXi$ is a continuum analogue of the particle numbering; 
  as a matter of fact, 
  the 1D label variable
  is a mere continuous interpolation 
  of the consecutive numbering of the particles, 
  but here we introduce the label variable 
  in a more abstract way 
  so as to enable extension to higher dimensions 
  without postulating regular numbering at all.
The moving curvilinear coordinate system 
  given by Eq.~(\ref{L.map})
  is sometimes referred to as 
  the \emph{convected coordinate system}\cite{Bird.Book1987}.
The description
  in which the field variables are regarded 
  as functions of $(\BoldXi,t)$ 
  is termed as the \emph{Lagrangian description}, 
  as opposed to the Eulerian description 
  in which the independent variables are $\mb{r}$ and $t$.

There are 
  infinitely many choices of label variable in general.
In the 1D case\cite{Ooshida.JPSJ80,Ooshida.PRE88}, 
  we choose $\xi$ so as to satisfy    
\begin{equation}
  {\dx}\xi({x},{t}) = \rho, \quad
  {\dt}\xi({x},{t}) = -Q
  \label{L3+4}, 
\end{equation}
  where $\rho = \rho(x,t)$ and $Q = Q(x,t)$
  are the density and the flux 
  satisfying the continuity equation (\ref{cont1D}).
It is then readily verified 
  that $\xi = \xi(x,t)$ 
  satisfies the 1D convective equation, 
  which qualifies $\xi$ as a label variable.

For this particular choice of the label variable, 
  the derivatives of $x = x(\xi,t)$ are 
\begin{equation}
  \dXi{x(\xi,t)} = \frac{1}{\rho(\xi,t)}, \quad 
  \dt {x(\xi,t)} = \frac{Q}{\rho} = u(\xi,t)
  \label{Dx}.
\end{equation}
The presence of $1/\rho$ in Eq.~(\ref{Dx})
  motivates us 
  to introduce the ``vacancy field'' 
\begin{equation}
  \psi = \psi(\xi,t) = \frac{\rho_0}{\rho(\xi,t)} - 1  
  \label{psi=},
\end{equation}
  so that the relation $\dt\dXi{x} = \dXi\dt{x}$  
  is cast into the form of the vacancy conservation, 
\begin{equation}
  \ell_0 \dt\psi(\xi,t) = \dXi{u(\xi,{t})}
  \label{cont.psi}.
\end{equation}


Then, by using $\psi(\xi,t)$ instead of $\rho(x,t)$, 
  we can derive a more rigorous version 
  of the AP formula\cite{Ooshida.PRE88},
  in the sense 
  that the approximation in Eq.~(\ref{AP.Eq2})
  is not required. 
The formula for the MSD reads
\begin{equation}
    \Av{R^2}  
    = \frac{L^4}{\pi N^2}  \int_{-\infty}^{\infty} 
    \frac{C(k,0) - C(k,t)}{k^2}
    \D{k} 
    \label{AP}
\end{equation}
  in the statistically steady case, 
  where 
\begin{equation}
  C(k,t) 
  \eqdef \frac{N}{L^2} \Av{\check\psi(k,t)\check\psi(-k,0)}
  \label{Corr=}
\end{equation}
  is the Lagrangian correlation of vacancy 
  in Fourier representation:  
\begin{equation}
  \check\psi(k,t) 
  = {\frac1N} \int e^{\II k\xi} \psi(\xi,t) \D\xi
  \relax, \quad
  \psi(\xi,t) = \sum_k \check\psi(k,t) e^{-\I k\xi}
  \quad
  \left(  \frac{k}{2\pi/N} \in \Z  \right)
  \label{m1+2}. 
\end{equation}
The ha\v{c}ek (inverted hat) is used
  to make a clear distinction between the two Fourier transforms 
  exemplified by Eqs.~(\ref{k1+3}) and (\ref{m1+2}).
The formula for the MSD in Eq.~(\ref{AP})
  is a special case 
  of the formula for the displacement correlation, 
\begin{equation}
  \Av{R(\xi,t)R(\xi',t)}  
  = \frac{L^4}{\pi N^2}  \int_{-\infty}^{\infty} 
  e^{-{\I}k(\xi-\xi')}  \frac{C(k,0) - C(k,t)}{k^2}
  \D{k} 
  \label{AP+}, 
\end{equation}%
  where $R(\xi,t) = x(\xi,t) - x(\xi,0)$.
For $\xi=\xi'$, Eq.~(\ref{AP+}) reduces to Eq.~(\ref{AP}). 

To begin the derivation of the AP formula,
  we notice that  
\begin{equation}
  \dXi{R(\xi,t,s)} 
  = \ell_0 \left[ \psi(\xi,t) - \psi(\xi,s) \right]
  \label{dR/dX}
\end{equation}
  where $R(\xi,t,s) = x(\xi,t) - x(\xi,s)$;
  this is readily verified 
  with Eqs.~(\ref{Dx}) and (\ref{psi=}). 
Integration of Eq.~(\ref{dR/dX}) 
  in Fourier representation 
  yields
\begin{equation}%
  R(\xi,t,s) 
  = 
  \ell_0 \sum_{k} 
  \left[ \check\psi(k,t) - \check\psi(k,s) \right]
  \frac{e^{-\I{k\xi}}}{-\I{k}}
  + \RG(t,s)
  \label{R//psi},
\end{equation}
  with $\RG$
  denoting the displacement of the center of mass, 
  negligible for $N\to\infty$.
Multiplying Eq.~(\ref{R//psi})  
  by its duplicate with $(\xi,k)$ replaced by $(\xi',-k')$, 
  we find  
\begin{align}
  &\Av{{R(\xi,t,s)}{R(\xi',t,s)}} \notag \\ 
  &= 
  \ell_0^2
  \sum_k \sum_{k'} 
  \Av{%
  \left[ \check\psi( k, t) - \check\psi( k, s) \right]
  \left[ \check\psi(-k',t) - \check\psi(-k',s) \right]}
  \frac{e^{-\I{k\xi}+\I{k'\xi'}}}{{k}{k'}}
  \notag \\
  &= 
  \ell_0^2
  \sum_k  
  \Av{%
  \left[ \check\psi( k,t) - \check\psi( k,s) \right]
  \left[ \check\psi(-k,t) - \check\psi(-k,s) \right]}
  \frac{e^{-\I{k(\xi-\xi')}}}{k^2}
  \label{RR.sum}\relax;  
\end{align}
  the terms with $k \ne {k'}$ is shown to vanish 
  if $\psi(k,t)$ is subject to linear Langevin equation
  whose forcing term is uncorrelated for different wave\-numbers
  [see Eq.~(\ref{f4}), for example], 
  and the same is true for nonlinear cases, 
  in the limit of $N\to\infty$,  
  as long as the vertex is sparse 
  in Kraichnan's sense%
  \cite{Kraichnan.JFM5,Goto.PhysicaD117}.  
Taking continuum limit 
  as 
  $\sum_{k}(\cdots)
  \to \frac{N}{2\pi}\int(\cdots)\D{k}$, 
  we obtain 
\begin{multline}
  \Av{{R(\xi,t,s)}{R(\xi',t,s)}}  \\ {}
  = 
  \frac{L^4}{\pi{N^2}}
  \int_{-\infty}^{+\infty}
  \left[ 
  \frac{C(k,t,t) + C(k,s,s)}{2} - C(k,t,s)
  \right]
  \frac{e^{-\I{k}(\xi-\xi')}}{k^2}  \D{k}
  \label{AP+.2time} 
\end{multline}
  where
  $ C(k,t,s) \eqdef 
  (N/L^2) \Av{\check\psi(k,t)\check\psi(-k,s)}
  $. 
In the statistically steady case,  
  we can choose $s=0$ without loss of generality
  and, setting $C(k,t,0) = C(k,t)$,
  which also implies that $C(k,t,t) = C(k,s,s) = C(k,0)$, 
  we arrive at Eq.~(\ref{AP+}).

The AP formula (\ref{AP+})  
  has the advantage 
  of being able to provide the displacement correlation, 
  which is a four-point space-time correlation,  
  in terms of two-body Lagrangian correlation.
This is contrastive 
  to the approach via $\Fs$ in Eq.~(\ref{Fs=}),  
  which requires calculation of a four-body correlation. 

\subsection{AP formula in higher dimensions}

The AP formula 
  may seem to be easily extensible to higher-dimensional cases, 
  simply with the $\nd$-dimensional wavenumber integral 
  as in Eq.~(\ref{EW.nD}). 
This form, however, should not be taken literally 
  but only as a simplified treatment 
  in which the deformation is restricted 
  to one of the $\nd$ modes of elastic waves%
  \cite{Toninelli.PRE71}. 
As was noted in Subsec.~\ref{subsec:elastic}, 
  the description of colloidal liquids 
  requires 
  distinction between the longitudinal and transverse modes.
Here we demonstrate 
  how to extend the AP formula respecting this distinction, 
  which results in a tensorial formula.  


For concreteness, 
  let us focus on the 2D case ($\nd=2$).
Our target 
  is the \emph{displacement correlation tensor}, 
\begin{equation}
  \Av{\mb{R}(\BoldXi,t,s)\otimes\mb{R}(\BoldXi',t,s)}
  = 
  \begin{bmatrix}
    \Av{{R_1}{R_1}} \!& \Av{{R_1}{R_2}} \\
    \Av{{R_2}{R_1}} \!& \Av{{R_2}{R_2}}
  \end{bmatrix}
  \label{RR=}
\end{equation}
  where $R_1$ and $R_2$ 
  are the Cartesian components of the displacement $\mb{R}$,
  with the arguments omitted when obvious.
The label variable $\BoldXi = (\xi,\eta)$ 
  is defined 
  through the 2D extension of Eq.~(\ref{L3+4}),  
  later shown as Eq.~(\ref{2D.L}),  
  so that $\bm{\Xi}_i = \BoldXi(\mb{r}_i(t),t)$ 
  is independent of $t$.
Conceptually, 
  $\ell_0\bm{\Xi}_i$ is supposed to express
  the mean position of the $i$-th particle
  (for some appropriate time\-scale 
  shorter than the $\alpha$ relaxation time\cite{Brito.EPL76}).

The ambiguity  
  due to non-uniqueness of $\BoldXi$
  can be removed 
  by introducing  
\begin{equation}
  \ChiR(\di,t,s)
  = \Av{\frac{1}{L^2}\iint
  \delta^2( \rB - \rA - \di )\, 
  \RA\otimes\RB\,
  {\D^2\rA}{\D^2\rB}} 
  \label{ChiR=}
\end{equation}
  where 
  $\rA = \mb{r}(\BoldXi_{\mathrm{A}},s)$,
  $\RA 
  = \mb{r}(\BoldXi_{\mathrm{A}},t) 
  - \mb{r}(\BoldXi_{\mathrm{A}},s)$, 
  and 
  $\rB$ and $\RB$
  are defined analogously.  
Equation (\ref{ChiR=}) 
  gives the two-particle displacement correlation 
  as a function 
  of the initial relative position vector $\di$.
Approximately   
  we have 
\begin{equation}
  \ChiR(\di,t,s) \simeq 
  \Av{\mb{R}(\BoldXi,t,s)\otimes\mb{R}(\BoldXi',t,s)}  
  \label{Chi.approx}
\end{equation}
  with $\di \simeq \ell_0(\BoldXi-\BoldXi')$, 
  and a correction to this approximate expression 
  can be obtained in terms of triple correlations
  (see Subsec.~V-A. in Ref.~\refcite{Ooshida.PRE88}).


Recalling 
  that our construction of the 1D label variable 
  through Eq.~(\ref{L3+4})
  is closely connected 
  to the 1D continuity equation (\ref{cont1D}), 
  we relate the $\nd$-dimensional label variable 
  to the $\nd$-dimensional continuity equation, 
\begin{equation}
  \dt\rho + \nabla\cdot\mb{Q} = 0  \label{+cont}, 
\end{equation}
  in a similar way.
The (particle-scale) density $\rho   = \rho(\mb{r},t) = \sum_i\rho_i$
  and the corresponding flux $\mb{Q} = \mb{Q}(\mb{r},t) = \rho\mb{u}$ 
  are given by straightforward generalization 
  of Eqs.~(\ref{rho=}) and (\ref{Q=}):
  for $\nd=2$, 
  we have 
  $\rho_i =  \delta^2({\mb{r}}-\mb{r}_i({t}))$
  and 
  $\mb{Q} = \sum_i \rho_i({\mb{r}},{t}) \dot{\mb{r}}_i({t})$.
  
To generalize Eq.~(\ref{L3+4}) 
  to $\nd$-dimensional cases, 
  we write it in the form 
\begin{equation}
  (\rho,Q) =
  \begin{vmatrix}
     \mb{e}_0^{} & {\partial_t}\xi \\
     \mb{e}_1^{} & {\partial_x}\xi 
  \end{vmatrix}
  = ({\dx\xi}, \;{-\dt\xi})
  \relax. 
\end{equation}
The above form is suggestive 
  and motivates us to assume 
  that $\BoldXi$
  is related to $(\rho,\mb{Q})$
  by the bilinear equation 
\begin{equation} 
  (\rho,\mb{Q}) =
  \begin{vmatrix}
      \mb{e}_0^{} & \dt\xi & \dt\eta \\
      \mb{e}_1^{} & \dx\xi & \dx\eta \\
      \mb{e}_2^{} & \dy\xi & \dy\eta 
  \end{vmatrix}
  =
  \left(
  \;
  \begin{vmatrix}
    \dx\xi & \dx\eta \\
    \dy\xi & \dy\eta 
  \end{vmatrix}, 
  \; 
  \begin{vmatrix}
    \dy        \xi & \dy        \eta \\
    {\dt}\xi & {\dt}\eta 
  \end{vmatrix}, 
  \; 
  \begin{vmatrix}
    {\dt}\xi & {\dt}\eta \\
    \dx        \xi & \dx        \eta 
  \end{vmatrix} 
  \;
  \right)
  \label{2D.L}
\end{equation}
  for $\nd=2$, 
  and a trilinear equation in an analogous form 
  for the 3D case\cite{Ooshida.JPSJ80}.
It is then verified 
  that $\BoldXi$ satisfies  
  the convective equation (\ref{convect}), 
  if it solves Eq.~(\ref{2D.L}).
Within linear approximation 
  in regard to $\rho-\rho_0$ and $\mb{Q}$, 
  a solution
\begin{equation}
  (\xi,\eta) 
  = \frac{(x,y)}{\ell_0} 
  - \ell_0\int_o^t\mb{Q}(\mb{r},\tilde{t})\,\D{\tilde{t}} 
  + \ell_0\nabla\phi, 
  \quad
  \nabla^2\phi = \rho(\mb{r},o) - \rho_0
  \label{xi.2D.approx}
\end{equation}
  can be constructed, 
  where $o$ is an arbitrary ``initial'' time.

In the spirit of Eqs.~(\ref{Dx}) and (\ref{psi=})
  where $\psi = \psi(\xi,t)$ is introduced 
  on the basis of $\partial{x}/\partial{\xi} = 1/\rho$, 
  we consider the deformation gradient tensor 
  and introduce $(\Psi_1,\Psi_2)$ 
  on the basis of its diagonal components, 
  as 
\begin{equation}
   \begin{bmatrix}
       {\dXi{x}} & {\dHa{x}}\\
       {\dXi{y}} & {\dHa{y}} 
   \end{bmatrix}
   = 
   \ell_0 \left( 
        \openone + 
        \begin{bmatrix}
            \Psi_1(\BoldXi,t) &  * \\ 
            *  &  \Psi_2(\BoldXi,t) 
        \end{bmatrix}
   \right)  
   \label{dx/dXi}, 
\end{equation}%
  where the off-diagonal components are omitted 
  and replaced with an asterisk.

It follows from Eq.~(\ref{dx/dXi}) 
  that 
  $\dXi{R_1}(\BoldXi,t,s) 
  = \ell_0 \left[ \Psi_1(\BoldXi,t) - \Psi_1(\BoldXi,s) \right]
  $, 
  with $s < t$, 
  and 
  an analogous expression for $\dHa{R_2}$ 
  is also readily available.
Introducing 
  the Fourier representation 
  as 
\begin{equation}
  \TwoVect{\Psi_1(\BoldXi,t)}{\Psi_2(\BoldXi,t)} 
  = \sum_{\mb{k}} 
  \TwoVect{\check\Psi_1(\mb{k},t)}{\check\Psi_2(\mb{k},t)} 
  e^{-\I\kXi}
  \label{2D.m2} 
\end{equation}
  with $\mb{k} = (k_1,k_2)$, 
  which is a straightforward generalization of Eq.~(\ref{m1+2}), 
  we find 
\begin{equation}
  R_1(\BoldXi,t,s) 
  = 
  \ell_0 \sum_{\mb{k}} 
  \left[ \check\Psi_1(\mb{k},t) - \check\Psi_1(\mb{k},s) \right] 
  \frac{e^{-\I\kXi}}{-\I{k_1}}
  + X_{\mathrm{G}}(t,s)
  \label{R1//psi};
\end{equation}
  subsequently, 
  with the fluctuation of the center of mass 
  $\mathbf{R}_{\mathrm{G}} = (X_{\mathrm{G}}, Y_{\mathrm{G}})$
  neglected for $N\to\infty$, 
  Eq.~(\ref{R1//psi}) yields 
\begin{align}
  &\Av{{R_1(\BoldXi,t,s)}{R_1(\BoldXi',t,s)}}  \notag\\ {}
  &= 
  \ell_0^2 \sum_{\mb{k}}\sum_{\mb{k}'}
  \Av{%
  \left[ \check\Psi_1( \mb{k}, t) - \check\Psi_1( \mb{k}, s) \right]
  \left[ \check\Psi_1(-\mb{k}',t) - \check\Psi_1(-\mb{k}',s) \right]
  }
  \frac{e^{-\I\kXi+\I\mb{k}'\cdot\BoldXi'}}{{k_1}{k'_1}}
  \relax.
\end{align}
Provided that 
  the same condition 
  as was mentioned immediately after Eq.~(\ref{RR.sum}) 
  is satisfied, 
  the terms with $\mb{k} \ne \mb{k}'$
  are seen to vanish as in the 1D case.  
Then, taking the continuum limit,
  we obtain an expression 
  that gives $\Av{{R_1}{R_1}}$
  in terms of correlation of $\check\Psi_1$.
Other components of the displacement correlation tensor 
  are calculated in a similar way.
Defining  
\begin{equation}
  C_{\alpha\beta}(\mb{k},t,s)
  \eqdef  
  \frac{N}{L^4} 
  \Av{{\check\Psi_\alpha(\mb{k},t)}{\check\Psi_\beta(-\mb{k},s)}}
  \quad
  (\alpha,\beta \in \{1,2\})
  \label{CorrTensor2=}   
\end{equation}
  and 
  assuming 
  $C_{\alpha\beta}$ 
  to be real and symmetric 
  so that $C_{21}(\pm\mb{k},t,s) = C_{12}(\mb{k},t,s)$,
  we find 
\begin{multline}
  \Av{{{R_\alpha}(\BoldXi,t,s)}{{R_\beta}(\BoldXi',t,s)}} \\ {}
  = \frac{L^6}{{2\pi^2}N}
  \iint
  \left[ 
  \frac{%
  C_{\alpha\beta}(\mb{k},s,s) + C_{\alpha\beta}(\mb{k},t,t)
  }{2}
  - C_{\alpha\beta}(\mb{k},t,s)
  \right]
  \frac{e^{-\I\mb{k}\cdot(\BoldXi-\BoldXi')}}{{k_\alpha}{k_\beta}}
  \D{k_1}\D{k_2}
  \label{AP2.ab}, 
\end{multline}
  where we can choose $\BoldXi' = \mb{0}$ 
  without loss of generality. 


For later convenience, 
  let us rewrite the 2D AP formula 
  in terms of the dilatational and rotational modes
  of deformation.
We define 
\begin{subequations}%
\begin{align}
  \psiD(\mb{k},t) 
  &= \check\Psi_1(\mb{k},t) + \check\Psi_2(\mb{k},t)
  \label{psiD=},
  \\
  \psiR(\mb{k},t) 
  &= \frac{k_1}{k_2} \check\Psi_1(\mb{k},t) 
  -  \frac{k_2}{k_1} \check\Psi_2(\mb{k},t)
  \label{psiR=}, 
\end{align}%
\end{subequations}%
  so as to have the Fourier representation 
  of the dilatational and rotational modes:
\begin{equation}
  \dXi{x} + \dHa{y} 
  = \ell_0
  \left( 2 + \sum_{\mb{k}} \psiD(\mb{k},t) e^{-\I\kXi} \right)
  \relax, 
  \quad   
  \dXi{y} - \dHa{x}
  = \ell_0 \sum_{\mb{k}} \psiR(\mb{k},t) e^{-\I\kXi} 
  \label{div/rot.X}.
\end{equation}
The correlations of these modes  
  are denoted by 
\begin{equation}
  C_{ab}(\mb{k},t,s)
  = 
  \frac{N}{L^4} 
  \Av{{\psi_a(\mb{k},t)}{\psi_b(-\mb{k},s)}}
  \label{CorrH=}
\end{equation}
  with $a, b \in \{\mathrm{d},\mathrm{r}\}$ and $s<t$. 
Obviously, 
  $C_{\alpha\beta}$ in Eq.~(\ref{CorrTensor2=}) 
  can be expressed as a linear combination 
  of $C_{ab}$ in Eq.~(\ref{CorrH=}), 
  and vice versa.
For simplicity,
  let us assume 
  that $\Crd$ and $\Cdr$ vanish identically 
  for some reason (such as the parity), 
  and denote $\Cdd$ by $\Cd$ and $\Crr$ by $\Cr$.
Then the relation between $C_{\alpha\beta}$ and $C_a$
  reads
\begin{equation}
  \begin{bmatrix}
    C_{11} & C_{12} \\ C_{21} & C_{22}
  \end{bmatrix}
  =
  \frac{1}{\mb{k}^4}  
  \begin{bmatrix}
    k_1^4 &  k_1^2 k_2^2 \\  k_1^2 k_2^2 &  k_2^4
  \end{bmatrix}
  \Cd
  +
  \frac{k_1^2 k_2^2}{\mb{k}^4}  
  \begin{bmatrix}
    1 & -1 \\ -1 & 1 
  \end{bmatrix}
  \Cr 
  \label{CorrH.LC};
\end{equation}
  note that $C_{\alpha\beta}$ is symmetric. 
Substituting $C_{\alpha\beta}$ in Eq.~(\ref{CorrH.LC})
  into Eqs.~(\ref{AP2.ab}) 
  and choosing $\BoldXi' = \mb{0}$ without loss of generality,
  we find 
\begin{align}
  &\relax
  \begin{bmatrix}
    \Av{{R_1(\BoldXi,t,s)}{R_1(\mb{0},t,s)}} &
    \Av{{R_1(\BoldXi,t,s)}{R_2(\mb{0},t,s)}} \\
    \Av{{R_2(\BoldXi,t,s)}{R_1(\mb{0},t,s)}} &
    \Av{{R_2(\BoldXi,t,s)}{R_2(\mb{0},t,s)}}   
  \end{bmatrix}
  \notag\\&{}
  = \frac{L^6}{{2\pi^2}N}
  \iint
  \left[ 
  \frac{\Cd(\mb{k},s,s) + \Cd(\mb{k},t,t)}{2} - \Cd(\mb{k},t,s)
  \right]
  \begin{bmatrix}
    k_1^2      & {k_1}{k_2} \\
    {k_2}{k_1} & k_2^2
  \end{bmatrix}
  \frac{e^{-\I\kXi}}{\mb{k}^4}  \D{k_1}\D{k_2}
  \notag\\&{}
  +  \frac{L^6}{{2\pi^2}N}
  \iint
  \left[ 
  \frac{\Cr(\mb{k},s,s) + \Cr(\mb{k},t,t)}{2} - \Cr(\mb{k},t,s)
  \right]
  \begin{bmatrix}
    k_2^2       & \!{-{k_1}{k_2}} \\
    -{k_2}{k_1} &      k_1^2
  \end{bmatrix}
  \frac{e^{-\I\kXi}}{\mb{k}^4}  \D{k_1}\D{k_2}
  \label{AP2.H}.
\end{align}
Equation (\ref{AP2.H}) 
  is our main result in this article.

As we will see in Sec.~\ref{sec:colloid.2D}, 
  Eq.~(\ref{AP2.H}) is simplified 
  in the isotropic case,  
  in which $\Cd$ and $\Cr$ are independent 
  of the directions of $\mb{k}$ 
  so that $C_a = C_a(k,t,s)$.
In this case, 
  with the approximation 
  $\di \simeq \ell_0(\BoldXi-\BoldXi')$,  
  the displacement correlation tensor 
  is expressible 
  in terms of two functions 
  $\Xl = \Xl(\xi,t)$ and $\Xtr = \Xtr(\xi,t)$: 
\begin{equation}
  \ChiR 
  = \Xl (\di[\relax]/\ell_0,t) \frac{\di\otimes\di}{\di^2}
  + \Xtr(\di[\relax]/\ell_0,t)
  \left( \openone - \frac{\di\otimes\di}{\di^2} \right)
  \label{Xl+Xtr}.
\end{equation}
The functions $\Xl$ and $\Xtr$ 
  denote the longitudinal and transverse 
  displacement correlations, respectively. 

\section{Displacement Correlation in 1D Colloidal Liquid}
\label{sec:colloid.1D}

To demonstrate 
  how to apply the AP formula to colloidal systems,  
  here we begin by illustrating the 1D problem.
Starting from the Dean--Kawasaki equation,  
  we calculate the Lagrangian correlation $C(k,t)$  
  and substitute it 
  into the 1D AP formula.
This procedure will be a prototype 
  for the 2D calculation 
  to be discussed in Sec.~\ref{sec:colloid.2D}.

\subsection{Formulation}

The dynamics of Brownian particles 
  in the over\-damped regime 
  can be expressed
  as a stochastic equation for the density field, 
  known as the Dean--Kawasaki equation%
  \cite{Dean.JPhAMG29,Kawasaki.PhysicaA208,Kawasaki.JStatPhys93}.
Utilizing the continuity equation (\ref{cont1D})
  that relates the density field $\rho(x,t)$
  to the flux $Q(x,t)$, 
  we write 
  the 1D Dean--Kawasaki equation 
  in the form 
\begin{equation}
  Q 
  = -D\,\left( \dx\rho + \frac{\rho}{\kT}\dx{U} \right)
  + \sum_j \rho_j(x,t) f_j(t) 
  \label{DK1.Q},  
\end{equation}
  where 
  $U = U[\rho](x) = \int{\Veff(x-x')\rho(x')}\D{x'}$ 
  describes the interaction of the particle, 
  with $\Veff$ denoting the effective potential 
  based on the direct correlation function.

Subsequently,  
  we rewrite Eq.~(\ref{DK1.Q}) 
  with the label variable $\xi$ 
  introduced through Eq.~(\ref{L3+4}).
The field variable is $\psi = \psi(\xi,t)$
  and the corresponding flux, 
  according to Eq.~(\ref{cont.psi}), 
  is $u = u(\xi,t)$.
Dividing both sides of Eq.~(\ref{DK1.Q}) by $\rho$ 
  and rewriting the differentials with the chain rule,  
  $\dx = \rho\dXi$, 
  we obtain 
\begin{align}
  u 
  &= 
  -D\left( \frac{\dx\rho}{\rho} + \frac{\dx{U}}{\kT} \right)
  + \sum_i \delta(\xi-\Xi_i) f_i(t) 
  \notag \\ 
  &= 
  -D\left( 
  \dXi\frac{\rho_0}{1+\psi} + {\frac{\rho}{\kT}}\dXi{U} 
  \right)
  + \dXi^{-1} \fL(\xi,t)\qquad
  \label{1D.Langevin.u}
\end{align}
  where 
  $\fL(\xi,t) 
  = \dXi \sum_i \delta\left( \xi - \Xi_i \right) f_i(t)
  $ 
  and 
  $\Xi_i = \xi(X_i,t)$.

Switching over to the Fourier representation  
  given in Eq.~(\ref{m1+2}), 
  we rewrite the combination 
  of Eqs.~(\ref{cont.psi}) and (\ref{1D.Langevin.u})
  as 
\begin{equation}
  \partial_t  \check\psi(k,t) 
    = 
    -\Dc_* k^2  
    \check\psi(k,t)  
    + \sum_{{p+q+k=0}} 
    {\Vertex_k^{pq}} \check\psi(-p,t) \check\psi(-q,t)
    + \rho_0 \CheckFL(k,t)
  \label{*3+},
\end{equation}
  with terms of $O(\check\psi^3)$ dropped. 
The concrete form of the vertex $\Vertex_k^{pq}$ 
  is given in Refs.~\refcite{Ooshida.JPSJ80} 
  and \refcite{Ooshida.PRE88}.
The statistics for the thermal forcing 
  are given by 
\begin{equation}
  \rho_0^2 \Av{\CheckFL(k,t) \CheckFL(-k',t')}
  = \frac{2 D_*}{N} k^2 \delta_{kk'} \delta(t-t')
  \label{f4}.
\end{equation}

\subsection{One-dimensional linear theory}
\label{subsec:colloid.1D.linear}

As an input into the AP formula, 
  let us calculate $C(k,t)$ 
  in Eq.~(\ref{Corr=})
  on the basis of linear approximation to Eq.~(\ref{*3+}).
We assume 
  that the system is in equilibrium,
  so that the initial condition 
  is not important.
Then the linearized equation 
  yields
\begin{equation}
  \check\psi(k,t) 
   = \int_{-\infty}^t \D{t'} 
   e^{-{\Dc_*} k^2 (t-t')}
   \rho_0 \CheckFL(k,t')
   \label{d3psi},
\end{equation}
  from which we obtain,  
  with the aid of Eq.~(\ref{f4}), 
\begin{equation} 
 C(k,t) 
 = \frac{S}{L^2} e^{-(D_*/S) k^2 t}, \quad
 S = S(k) \simeq S(0)  
 \label{d4psi}.
\end{equation}

Using the the AP formula (\ref{AP+}),
  we convert the vacancy correlation $C$ 
  into the displacement correlation, 
  $\Av{ R(\xi,t) R(\xi',t) }$.  
The result is shown to be expressible 
  in terms of the similarity variable 
  $\theta = {\ell_0(\xi-\xi')}/{\ld(t)}$,
  with 
  $\lambda(t) = 2\sqrt{{\Dc}t}$, 
  as 
\begin{equation}
  \Av{ R(\xi,t) R(\xi',t) }
  = 
  \frac{2S}{\rho_0} \sqrt{\frac{{\Dc}t}{\pi}}\,
  \left(
  {e^{-\theta^2} - \sqrt{\pi} \abs{\theta} \erfc\abs\theta}
  \right)
  \label{R*R}.
\end{equation}
Setting $\theta = 0$ in Eq.~(\ref{R*R})
  reproduces 
  the established HKK law (\ref{HKK}).
The prediction of Eq.~(\ref{R*R}) 
  is compared 
  with numerically computed 
  displacement correlations\cite{Ooshida@Rhodes1407+} 
  in Fig.~\ref{Fig:DC.1D}(b),
  with $j-i$ interpreted as $\xi-\xi'$.
The numerical correlation 
  for the higher barrier case ($\Vmax = 5\kT$) 
  is consistent with Eq.~(\ref{R*R}).


Incidentally,
  the analysis can be extended to a case of age\-ing.
Suppose 
  that the system has been equilibrated,  
  in the presence 
  of an extra repulsive interaction between the particles, 
  in the state characterized 
  by the static structure factor $\Sinit(k)$.
Subsequently,
  the extra interaction is switched off at $t=0$, 
  so that the system relaxes toward a new equilibrium state.
Using 
  the linear solution
\begin{equation}
  \check\psi(k,t) 
   = \check\psi(k,0) e^{-{\Dc_*} k^2 t}
   + \int_0^t \D{t'} e^{-{\Dc_*} k^2 (t-t')}
   \rho_0 \CheckFL(k,t')
   \label{ic0.psi}
\end{equation}
  and Eq.~(\ref{f4}), 
  we find 
\begin{equation}
  C(k,t,s) 
  = \frac{S}{L^2}          e^{-{\Dc_*} k^2 (t-s)}
  + \frac{\Sinit - S}{L^2} e^{-{\Dc_*} k^2 (t+s)}
  \label{ic0.C}
\end{equation}
  with $0<s<t$, $\Sinit \simeq \Sinit(0)$ and $S \simeq S(0)$.
Note that, as expected,
  Eq.~(\ref{ic0.C}) reduces to Eq.~(\ref{d4psi})
  for $\Sinit \to S$.

It is then straightforward 
  to calculate the MSD in this age\-ing case.
By substituting Eq.~(\ref{ic0.C}) 
  into Eq.~(\ref{AP+.2time}) with $\xi = \xi'$, 
  we obtain 
\begin{multline}
  \Av{\smash{\left[R(t,s)\right]^2}}
  = \frac{2S}{\rho_0} \sqrt{\frac{\Dc(t-s)}{\pi}} 
  \\ {}
  + \frac{\Sinit-S}{\rho_0} 
  \left(
  2\sqrt{\frac{\Dc(t+s)}{\pi}} 
  - \sqrt{\frac{{2\Dc}t}{\pi}}  
  - \sqrt{\frac{{2\Dc}s}{\pi}}\,  
  \right)
  \label{ic0.MSD}.
\end{multline}
In studying age\-ing effects\cite{Kob.PRL78}, 
  it is more convenient 
  to rewrite the time arguments in Eq.~(\ref{ic0.MSD}) 
  as $s=\Tw$ and $t=\Tw+t'$, 
  denoting the ``waiting time'' with $\Tw$.
It is then readily shown  
  that $\Av{\smash{\left[R(\Tw+t',\Tw)\right]^2}}/\sqrt{t'}$
  is a function of $t'/\Tw$ only. 
This function 
  is asymptotically independent of $\Sinit$
  for $t'\ll \Tw$, 
  so that the HKK law (\ref{HKK}) is recovered;
  on the other hand, 
  as $t'$ elapses and exceeds $\Tw$,
  the effect of $\Sinit$ reappears.
For simplicity,  
  let us focus on the case of the initial condition
  with equally spaced particles\cite{Leibovich.PRE88},  
  which corresponds to $\Sinit=0$.
For this case, 
  Eq.~(\ref{ic0.MSD}) yields 
\begin{equation}
  \frac{%
  \Av{\smash{\left[R(\Tw+t',\Tw)\right]^2}}}{%
  (2S/\rho_0) \sqrt{\Dc{t'}/\pi}}
  =
  \begin{cases}
    1                 &  (t' \ll \Tw) \\ 
    \frac{1}{\sqrt2}  &  (t' \gg \Tw) \relax,
  \end{cases}
  \label{MSD.Leibovich}
\end{equation}
  showing that, if $t'$ is the same,
  the ratio of the MSD for long $\Tw$
  to that for short $\Tw$ is $\sqrt{2}:1$.
Thus the present analysis, valid for $\sigma\ge0$,  
  gives the same factor $\sqrt{2}$ 
  as was predicted by Leibovich and Barkai%
  \cite{Leibovich.PRE88}
  for point particles ($\sigma=0$).

\subsection{Nonlinear analysis of 1D colloidal liquid}
\label{subsec:colloid.1D.NL}

Now we return to the statistically steady case 
  and discuss the effect of the nonlinearity in Eq.~(\ref{*3+})
  which was ignored in Subsec.~\ref{subsec:colloid.1D.linear}.
There are two sources of nonlinearity: 
  the one originating from the direct contact between particles 
  is unimportant for low density 
  so it may suffice to incorporate it via $\Dc = D/S$, 
  but there is another kind of nonlinearity 
  that comes from the configurational entropy, 
  whose effect is not negligible even in the case of low density.

The nonlinearity of the configurational entropy 
  manifests itself 
  as a memory effect due to vacancy--vacancy interaction%
  \cite{Ooshida.PRE88}. 
For SFD on a lattice,  
  this memory effect   
  was pointed out 
  by van Beijeren \textit{et al.}\cite{van-Beijeren.PRB28}\relax; 
  assuming some phenomenological rules 
  about the dynamics of a vacancy cluster, 
  they calculated $\Av{R^2}$ 
  for both short and long $t$.
The crossover between the two limiting cases 
  of ${D_*}t \ll 1$ and ${D_*}t \gg 1$ 
  in the case of continuous systems 
  was taken into account, 
  again phenomenologically,  
  in Eq.~(\ref{MSD.Rallison})
  by Rallison\cite{Rallison.JFM186}
  and in Eq.~(\ref{MSD.Taloni}) 
  by Taloni and Lomholt\cite{Taloni.PRE78}.


It is then interesting to ask 
  whether a more systematic treatment of the nonlinearity
  can improve the phenomenological results 
  in Eqs.~(\ref{MSD.Rallison}) and (\ref{MSD.Taloni}).
Multiplying Eq.~(\ref{*3+}) with $\check\psi(-k,t)$
  and developing mode-coupling theory 
  to evaluate the triple correlation,  
  we find 
  the Lagrangian correlation $C$ to be governed by   
\begin{equation}
  \left( \dt + {\Dc_*}k^2 \right) C(k,t)  
  = -\int_0^t \ML(k,t-t') \partial_{t'} C(k,t') \D{t'}
  \label{L-MCT}
\end{equation}
  with the memory function $\ML$ 
  which is a quadratic functional of $C$.
By calculating  $C$ from Eq.~(\ref{L-MCT}) 
  and substituting it into the AP formula (\ref{AP}), 
  our group\cite{Ooshida.PRE88} found
\begin{equation}
  \Av{R^2}
  = \frac{2}{\rho_0} \sqrt{\frac{D t}{\pi}} 
  - \frac{\sqrt2}{3\pi} \rho_0^{-2}
  \label{MSD.L-MCT}.
\end{equation}

\begin{figure}
  \centering
  \raisebox{0.27\linewidth}{(a)}\hspace{-0.5em}%
  \IncFig[width=0.45\linewidth]{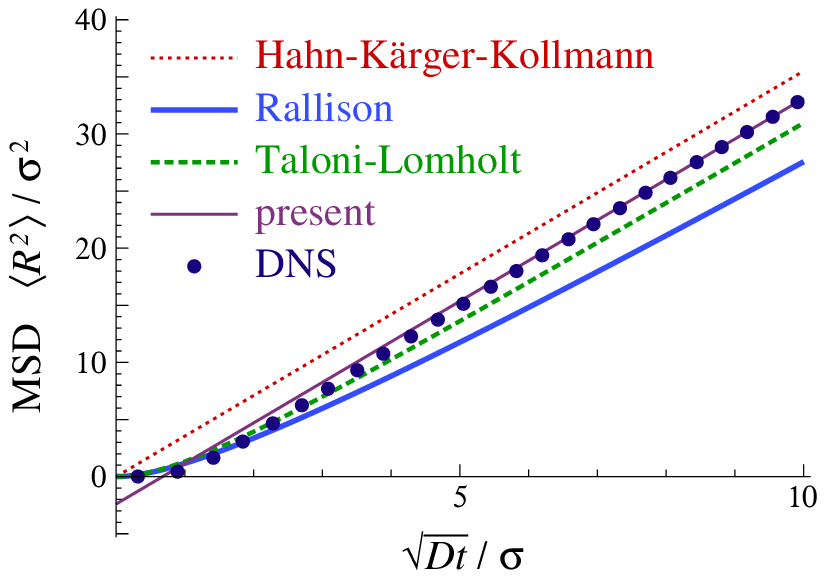}%
  \quad
  \raisebox{0.27\linewidth}{(b)}\hspace{-0.5em}%
  \IncFig[width=0.45\linewidth]{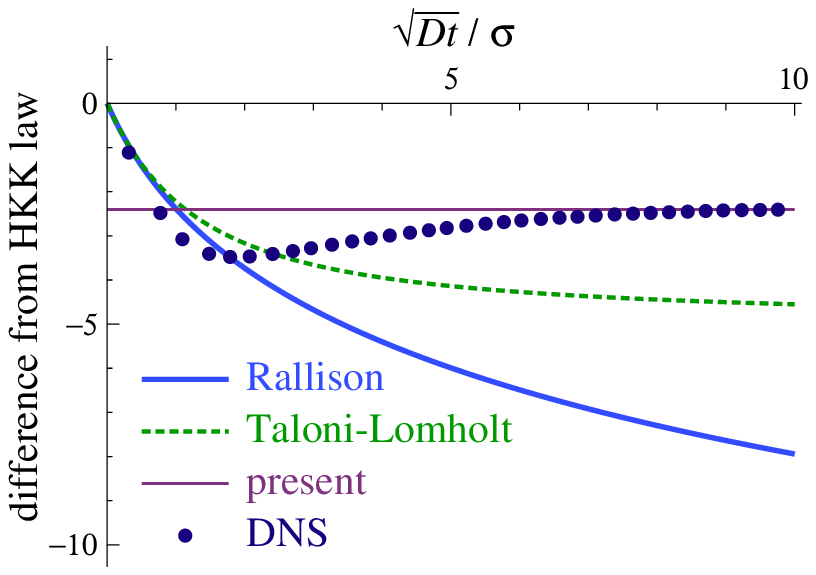}%
  \caption{\label{Fig:MSD1306}%
     Comparison of the 1D MSD 
     calculated 
     by the HKK law (\ref{HKK}), 
     Rallison's equation (\ref{MSD.Rallison}), 
     the Taloni--Lomholt solution (\ref{MSD.Taloni}),
     and the present result in Eq.~(\ref{MSD.L-MCT}), 
     with the result of DNS\protect\cite{Ooshida.PRE88}
     where   
     $N = 2^{15}$ and $\rho_0 = N/L = 0.25/\sigma$.
     (a) 
     The values of MSD, 
     nondimensionalized with $\sigma^2$, 
     are plotted against $\sqrt{Dt}/\sigma$.
     The straight line shows 
     the HKK law, $\Av{\smash{R^2}}_{\text{HKK}}$, 
     in Eq.~(\ref{HKK}).
     (b)
     The difference between the calculated MSD 
     and $\Av{\smash{R^2}}_{\text{HKK}}$
     for each case  
     are plotted in the same way as in (a).
  }
\end{figure}

In Fig.~\ref{Fig:MSD1306}, 
  the MSD in Eq.~(\ref{MSD.L-MCT})  
  is compared
  with direct numerical simulation (DNS) 
  of interacting Brownian particles. 
The phenomenological equations 
  (\ref{MSD.Rallison}) and (\ref{MSD.Taloni})
  are also included.  
Since $S = 0.624$ in this case  
  differs considerably from unity,
  we have revived $S$ in the equations in an obvious way 
  to make the asymptotic behavior 
  consistent with the HKK law (\ref{HKK}).

With $\sqrt{Dt}/\sigma$ taken as the horizontal axis,  
  the HKK law (\ref{HKK}) 
  is expressed as a straight line 
  passing through the origin 
  in Fig~\ref{Fig:MSD1306}(a).
Comparing this straight line ($\Av{R^2}_{\text{HKK}}$)
  with the result of DNS\cite{Ooshida.PRE88}
  ($\Av{R^2}_{\text{DNS}}$, bullets),  
  we notice that a negative correction is needed.
Judging from the graph 
  of $\Av{R^2}_{\text{DNS}} - \Av{R^2}_{\text{HKK}}$ 
  in Fig~\ref{Fig:MSD1306}(b),  
  the correction should become a negative constant 
  for $t\to\infty$.

All the three equations,
  namely Eqs.~(\ref{MSD.Rallison}), (\ref{MSD.Taloni}),
  and (\ref{MSD.L-MCT}),
  propose 
  negative correction to Eq.~(\ref{HKK}).
Figure~\ref{Fig:MSD1306} 
  allows comparing the predictions of these three equations 
  with the result of DNS
  (except for the close vicinity of $t=0$ 
  governed by the inertial effect). 
The logarithmic term in Eq.~(\ref{MSD.Rallison})
  is found to give an overcorrection.
The behavior of Eq.~(\ref{MSD.Taloni}) 
  seems close to that of DNS 
  for $\sqrt{Dt} < \ell_0 = 4\sigma$,  
  but for larger $t$ it also becomes over\-corrective.
Among the three equations, 
  the best estimation of the asymptotic behavior 
  is given by Eq.~(\ref{MSD.L-MCT}),
  which emphasizes the usefulness 
  of the AP formula combined with Lagrangian MCT.
 
\section{2D Colloidal Liquid}
\label{sec:colloid.2D}

Finally we have arrived at the stage 
  where analysis of 2D colloidal liquids  
  is within our reach.
The greatest difference 
  in comparison to the 1D case 
  is the presence of the transverse mode.

\subsection{Linear analysis of 2D colloidal liquid}

Here we target on calculation 
  of both the longitudinal and transverse 
  displacement correlations, 
  namely $\Xl$ and $\Xtr$ in Eq.~(\ref{Xl+Xtr}),
  based on linear analysis 
  of the 2D Dean--Kawasaki equation. 

The approach 
  is basically in parallel with the 1D case.
We start from the Dean--Kawasaki equation, 
  written in the form 
\begin{equation}
  \mb{Q} 
  = -D\,\left( \nabla\rho + \frac{\rho}{\kT}\nabla{U} \right)
  + \sum_j \rho_j(\mb{r},t) \mb{f}_j(t) 
  \label{Langevin.Q}
\end{equation}
  where 
  $ 
  U = U[\rho](\mb{r}) 
  = \int \Veff(r_*)\rho(\tilde{\mb{r}})\D^2\tilde{\mb{r}} 
  $ 
  with $r_* = \left|{\mb{r} - \tilde{\mb{r}}}\right|$.  
The thermal noise $\mb{f}_j(t)$ 
  is subject to Eq.~(\ref{fvect}).
Changing the independent variables 
  from $(\mb{r},t)$ to $(\BoldXi,t)$ 
  and taking the field variable $(\Psi_1,\Psi_2)$
  whose conservation law is given as
\begin{equation}
  \ell_0 \dt 
  \TwoVect
  {\Psi_1(\BoldXi,t)}
  {\Psi_2(\BoldXi,t)}
  = 
  \TwoVect
  {\dXi{u_x(\BoldXi,t)}}
  {\dHa{u_y(\BoldXi,t)}}
  \label{cont.Psi}, 
\end{equation}
  we rewrite Eq.~(\ref{Langevin.Q})
  in the form corresponding to Eq.~(\ref{1D.Langevin.u}): 
\begin{equation}
  \mb{u} 
  = -D
  \left ( 
  \frac{\nabla\rho}{\rho} + \frac{\nabla{U}}{\kT} 
  \right)
  + \sum_j \delta^2(\BoldXi-\bm{\Xi}_j) \mb{f}_j(t) 
  \label{Langevin.u},
\end{equation} 
  where
  $\nabla = (\nabla\xi) \dXi + (\nabla\eta) \dHa$ 
  is expressible in terms of $\Psi_1$ and $\Psi_2$.


Linear analysis of Eqs.~(\ref{cont.Psi}) and (\ref{Langevin.u}) 
  is then performed.
In terms of the Fourier modes 
  introduced in Eq.~(\ref{2D.m2}),
  the linearized equation reads
\begin{equation}
  \dt
  \TwoVect{\check\Psi_1(\mb{k},t)}{\check\Psi_2(\mb{k},t)}
  = 
  -\Dc_*
  \begin{bmatrix}
    k_1^2 & k_1^2  \\[0.5ex]%
    k_2^2 & k_2^2  
  \end{bmatrix}%
  \TwoVect{\check\Psi_1(\mb{k},t)}{\check\Psi_2(\mb{k},t)} 
  + 
  \ell_0^{-1} 
  \TwoVect{\check{f}_1(\mb{k},t)}{\check{f}_2(\mb{k},t)}
  \label{2D-lin.Psi.eqn}, 
\end{equation}
  with the forcing statistics given by
\begin{equation}
  \ell_0^{-2}
  \begin{bmatrix}
    \Av{{\check{f}_1(\mb{k},t)}{\check{f}_1(-\mb{k}',t')}}&
    \Av{{\check{f}_1(\mb{k},t)}{\check{f}_2(-\mb{k}',t')}}\\
    \Av{{\check{f}_2(\mb{k},t)}{\check{f}_1(-\mb{k}',t')}}&
    \Av{{\check{f}_2(\mb{k},t)}{\check{f}_2(-\mb{k}',t')}}    
  \end{bmatrix}
  = 
  \frac{2{D_*}}{N}
  \begin{bmatrix}
    k_1^2 & 0     \\ 
    0     & k_2^2 
  \end{bmatrix}
  \delta_{\mb{k},\mb{k}'} \delta(t-t')
  \label{f.2D}, 
\end{equation}
  where the delta-correlation in regard to the wavenumber 
  originates from the randomness of the particle configuration;
  possible correction at shorter length scales 
  is ignored for the present.

The linear part of Eq.~(\ref{2D-lin.Psi.eqn})  
  has two eigenvalues: 
  $k_1^2 + k_2^2$ and $0$, 
  correspond to the dilatational and rotational modes,
  respectively.
The correlations 
  $C_{\alpha\beta}(\mb{k},t,s)$ in Eq.~(\ref{CorrTensor2=}) 
  are then calculated, 
  according to Eq.~(\ref{CorrH.LC}), 
  as linear combinations 
  of $\Cd = \Cd(\mb{k},t,s)$ and $\Cr = \Cr(\mb{k},t,s)$. 
These correlations 
  are given as 
\begin{subequations}%
\begin{align}
  \Cd(\mb{k},t,s)
  &= 
  \frac{S}{L^4} \exp\left[ -\Dc_*\mb{k}^2(t-s) \right]
  \label{Lin2.Cd},
  \\
  \Cr(\mb{k},t,s)
  &= 
  \frac{2D_*\mb{k}^2}{L^4} (s - o)
  \label{Lin2.Cr},
\end{align}%
\label{eqs:Lin2}%
\end{subequations}%
  where $o$ is a constant of integration, 
  interpretable as the time at which $\Cr$ is reset. 
Note that both $\Cd$ and $\Cr$ are independent 
  of the direction of $\mb{k}$.


The last step 
  is to convert $\Cd$ and $\Cr$ 
  into the displacement correlation.
Using 
  the 2D AP formula 
  in the form of Eq.~(\ref{AP2.H}) 
  and introducing 
  $\theta^2 = (\xi^2 + \eta^2) / (4{\Dc_*}t) \simeq \di^2/[\ld(t)]^2$
  with $\ld(t) = 2\sqrt{{\Dc}t}$,    
  we obtain 
\begin{subequations}%
\begin{align}
 &\Av{{R_1(\BoldXi,t,s)}{R_1(\mb{0},t,s)}}
  = 
  \frac{S}{{4\pi}\rho_0}
  \left[ 
  E_1(\theta^2) 
  + \frac{\xi^2-\eta^2}{\xi^2+\eta^2} \frac{e^{-\theta^2}}{\theta^2} 
  \right]
  \label{Lin2.R1*R1}
  \\
 &\Av{{R_2(\BoldXi,t,s)}{R_2(\mb{0},t,s)}}
  = 
  \frac{S}{{4\pi}\rho_0}
  \left[ 
  E_1(\theta^2) 
  + \frac{-\xi^2+\eta^2}{\xi^2+\eta^2} \frac{e^{-\theta^2}}{\theta^2} 
  \right]
  \label{Lin2.R2*R2}
  \\
 &\Av{{R_1(\BoldXi,t,s)}{R_2(\mb{0},t,s)}}
  = 
  \frac{S}{{2\pi}\rho_0}
  \frac{\xi\eta}{\xi^2+\eta^2} \frac{e^{-\theta^2}}{\theta^2} 
  \label{Lin2.R1*R2}, 
\end{align}%
\label{eqs:Lin2.R*R}%
\end{subequations}%
  which can be rearranged, 
  in terms of $\phi$ such that 
  $\di = (\di[\relax]\,{\cos\phi},\;\di[\relax]\,{\sin\phi})$, 
  as 
\begin{equation}
  \Av{{\mb{R}(\BoldXi,t,s)}\otimes{\mb{R}(\mb{0},t,s)}}
  = 
  \frac{S}{{4\pi}\rho_0}
  \begin{bmatrix}
    E_1(\theta^2) 
    + \dfrac{e^{-\theta^2}}{\theta^2}\cos{2\phi} &
    \dfrac{e^{-\theta^2}}{\theta^2}\sin{2\phi} \\
    \dfrac{e^{-\theta^2}}{\theta^2}\sin{2\phi} &
    E_1(\theta^2) 
    - \dfrac{e^{-\theta^2}}{\theta^2}\cos{2\phi}  
  \end{bmatrix} 
  \label{Lin2.RR}
  \relax.
\end{equation}
By comparing Eq.~(\ref{Lin2.RR}) with Eq.~(\ref{Xl+Xtr}),
  on the ground that the expression 
  on the left-hand side of Eq.~(\ref{Lin2.RR}) 
  equals $\ChiR$ according to Eq.~(\ref{Chi.approx}),
  we reach the goal 
  of forming expressions for $\Xl$ and $\Xtr$. 
For large $\theta$,  
  using the asymptotic form 
  of the exponential integral $E_1(\theta^2)$,
  we find  
\begin{subequations}%
\begin{align}
  \Xl(\xi,t) 
  &= 
  \frac{S}{{4\pi}\rho_0}
  \left[
  E_1({\theta^2}) + \frac{e^{-\theta^2}}{\theta^2}
  \right]
  \simeq 
  \frac{S}{{4\pi}\rho_0} e^{-\theta^2}
  \left( 2 \theta^{-2} - \theta^{-4} + \cdots \right)
  \label{Xl} 
  \\
  \Xtr(\xi,t) 
  &= 
  \frac{S}{{4\pi}\rho_0}
  \left[
  E_1({\theta^2}) - \frac{e^{-\theta^2}}{\theta^2}
  \right]
  \simeq 
  \frac{S}{{4\pi}\rho_0} e^{-\theta^2}
  \left( - \theta^{-4} + \cdots \right)
  \label{Xtr}
  \relax.
\end{align}%
\label{eqs:ChiR.L}%
\end{subequations}%

\subsection{Discussion}

According to Eq.~(\ref{Xl}),
  the longitudinal displacement correlation $\Xl$ is positive, 
  while the transverse displacement correlation $\Xtr$ 
  in Eq.~(\ref{Xtr})
  is negative.
This is qualitatively consistent 
  with the vortical cooperative motion 
  observed by Doliwa and Heuer\cite{Doliwa.PRE61}.
Numerical verification of Eqs.~(\ref{eqs:ChiR.L})   
  will be reported elsewhere\cite{collab3.RR@PRE}.

By analogy 
  with the 1D displacement correlation in Eq.~(\ref{R*R})
  whose limiting value for $\theta \to 0$ correctly gives 
  the HKK law (\ref{HKK}) of MSD, 
  one may naturally expect 
  that the 2D displacement correlation 
  in Eqs.~(\ref{eqs:ChiR.L}) 
  will provide the MSD in the limit of $\theta\to0$.
Unfortunately, 
  it turns out that the 2D linear theory does not give 
  a quantitatively correct value of the MSD, 
  because it is not accurate enough in the short-wave region.
We may still attempt, however,  
  a qualitative estimation of the MSD 
  in terms of $\Xl$ in Eq.~(\ref{Xl}).
Introducing 
  the nondimensionalized cutoff length 
  $\xi_{\text{cut}} \sim \sigma/\ell_0$ 
  and using asymptotic form of the exponential integral 
  for small $\theta$, 
  we may estimate the MSD as 
\begin{equation} 
  \Av{\smash{\mb{R}^2}} 
  \sim \Xl(\xi_{\text{cut}},t)
  \sim 
  {O(1)}\times\frac{Dt}{{\rho_0}\sigma^2}
  + \frac{O(1)}{\rho_0}\log{\frac{Dt}{\sigma^2}}
  \label{MSD.2D.estimated}.
\end{equation}
Thus we obtain normal diffusion 
  with a logarithmic correction term. 
This seems to be qualitatively consistent 
  with the result 
  of van Beijeren and Kutner\cite{van-Beijeren.PRL55}
  for 2D lattices.
The logarithmic correction 
  seems to have the same physical origin 
  as the MSD for 2D elastic bodies in Eq.~(\ref{MSD.2D}).

Nevertheless, 
  we must emphasize 
  that the validity of Eq.~(\ref{MSD.2D.estimated}) 
  is rather limited   
  and, in particular, 
  it is definitely incorrect 
  for dense colloidal systems 
  such as the one depicted in Fig.~\ref{Fig:DC.1D}(a).
This is not the fault of the AP formula
  but the limitation of the linear theory,  
  as the time needed for the $\alpha$ relaxation 
  is largely underestimated 
  by ignoring the nonlinear effects 
  of the interacting particles
  with finite $\sigma$.
Evidently, 
  a nonlinear theory for $\Cd$ and $\Cr$ 
  should be developed and used instead of Eqs.~(\ref{eqs:Lin2}).
Combination of such nonlinear theory and the AP formula 
  will provide a better estimation of the MSD in 2D systems.
  
Some insights into the $\alpha$ relaxation
  may be given
  by comparative discussion 
  on SFD in quasi-1D systems and 2D dense colloidal liquid.
The $\alpha$ relaxation in colloidal liquid
  is believed to involve cooperation of numerous particles
  that occurs in some localized yet extended mode,   
  often termed 
  as a cooperatively rearranging region\cite{Adam.JCP43}. 
As a drastic simplification 
  of a cooperatively rearranging region
  in the spirit of mean-field approximation, 
  we may replace most of the particles with confining walls 
  and consider only two particles explicitly, 
  so that the $\alpha$ relaxation is modeled 
  by the positional exchange of two particles 
  in a quasi-1D channel.
Let us denote the position vector of the particles
  with $\rvect_1 = (X_1,Y_1)$ and $\rvect_2 = (X_2,Y_2)$, 
  and the potential with 
\begin{equation}
  U = U(\mb{r}_1,\mb{r}_2) 
  = V(r_{12}) + \Uex(Y_1) + \Uex(Y_2)
  \label{U12}
\end{equation}
  where $V(r_{12})$ is the interaction potential 
  as a function of $r_{12} = \abs{\rvect_2 - \rvect_1}$
  and $\Uex(Y_i)$ is the confinement potential 
  for the $i$-th particle.
The barrier potential 
  for the positional exchange of $X_1$ and $X_2$
  is given in terms of configurational integral,
  as  
\begin{equation}
  V_{\text{1D}} 
  = V_{\text{1D}}(X_2-X_1) 
  = -\kT\log\left( \iint e^{-{\beta}U}\D{Y_1}\D{Y_2} \right)
  + \text{const.}
  \label{V.1D},
\end{equation}
  where $\beta = 1/(\kT)$.
The ``hopping rate'' 
  for the positional exchange\cite{Wanasundara.JCP140}, 
  denoted with $1/\tauHop$, 
  is proportional to $e^{-\beta\max{V_{\text{1D}}}}$. 
Let us discuss how $\tauHop$ depends on $\Uex$, 
  comparing two cases:
\begin{subequations}%
\begin{align}
  \Uex(Y_i) &= U_{\text{soft}}(Y_i) = {\frac{\kappa}{2}}Y_i^2
  \label{U.soft}
\intertext{and}  
  \Uex(Y_i) &= U_{\text{hard}}(Y_i) 
  = 
  \begin{cases}
    0       & (\abs{Y_i} < L_y)  \\        
    \infty  & (\abs{Y_i} > L_y)  \relax. 
  \end{cases}
  \label{U.hard}
\end{align}%
\end{subequations}%
By assuming $V(r)$
  to be rigid sphere potential in both cases
  for the sake of simplicity, 
  the integral in Eq.~(\ref{V.1D}) can be concretely evaluated.
In the case of the soft wall in Eq.~(\ref{U.soft}), 
  we find $\log\tauHop \propto \kappa\sigma^2/(\kT)$
  in the limit of low temperature; 
  this expression of $\tauHop$ has no singularity 
  as a function of $\kappa$ and $\sigma$.
Contrastively, 
  $\tauHop$ in the hard wall case with Eq.~(\ref{U.hard}) 
  has a singularity as a function of $L_y$ and $\sigma$, 
  as it diverges for $\sigma/L_y \to 1/2$. 
Thus we find 
  that the $\sigma$-dependence of $\tauHop$, 
  or the dependence of $\tauHop$ 
  on the strength of the wall confinement,
  can be sharp or blunt 
  depending on whether the wall is rigid or not.
This behavior seems to be consistent 
  with the numerical findings 
  of Lucena \textit{et al.}\cite{Lucena.PRE85}
In the case of 2D and 3D colloidal liquids,
  the corresponding problem 
  is how the timescale of the $\alpha$ relaxation 
  depends on the volume fraction $\phi$.
It is known 
  that this timescale can exhibit 
  an extremely sharp dependence on $\phi$, 
  expressible as $e^{B/(\phi_0-\phi)}$ 
  with $B$ and $\phi_0$ being some constants%
  \cite{Reichman.JStat2005,Kawasaki.JPhys23};
  in comparison,  
  the singularity of 
  $\tauHop$ for $\Uex(Y_i) = U_{\text{hard}}(Y_i)$
  in the quasi-1D channel
  is much weaker, as it diverges only algebraically.
This difference suggests 
  subtlety of the cooperative dynamics 
  in the $\alpha$ relaxation in colloidal liquids
  that is difficult to capture 
  by the ``mean-field'' approach discussed here.

\section{Concluding remarks}
\label{sec:conc}

In search of insight into cooperativity 
  associated with the cage effect 
  in 
  glassy liquids,  
  we have developed a formalism 
  to calculate the displacement correlation tensor, 
  $\Av{\mb{R}(\BoldXi,t)\otimes\mb{R}(\BoldXi',t)}$,
  in $\nd$-dimensional systems.
The calculation 
  relies on the label variable method, 
  i.e.\ 
  adoption of the Lagrangian description%
  \cite{Ooshida.JPSJ80,Ooshida.PRE88}. 

In the 1D case, 
  Eq.~(\ref{AP}) gives the MSD
  as a special case 
  of the displacement correlation in Eq.~(\ref{AP+}). 
These equations 
  improve on the original formula (\ref{AP.original})  
  in that they are asymptotically exact for $N\to\infty$
  by virtue of the Lagrangian description. 
By combining Eq.~(\ref{AP})
  with the Lagrangian MCT equation (\ref{L-MCT}), 
  we have calculated 
  a finite-time correction 
  to the asymptotic HKK law (\ref{HKK}).
The formula was also applied to an age\-ing SFD 
  and reproduced the effect of the initial condition on the MSD, 
  previously obtained by Leibovich and Barkai\cite{Leibovich.PRE88}
  with a different approach.
It will be interesting 
  to study various extensions of SFD
  with the AP formula (\ref{AP}), 
  so that the effect 
  of spatially correlated noise\cite{Tkachenko.PRE82} 
  may be clarified, for example.

Our main result is the 2D formula (\ref{AP2.H}), 
  which is tensorial. 
For isotropic systems, 
  the displacement correlation tensor 
  reduces 
  to the longitudinal correlation $\Xl$
  and the transverse correlation $\Xtr$.
Linear analysis of the two-dimensional colloidal system 
  predicts $\Xtr > 0$ and $\Xl < 0$, 
  which seems to be qualitatively consistent 
  with the vortical cooperative motion 
  observed in numerical simulations\cite{Doliwa.PRE61}.

As a future direction,  
  it is promising 
  to improve the 2D result 
  by introducing nonlinear theory for $\Cd$ and $\Cr$, 
  for which 
  Eq.~(\ref{L-MCT}) provides an encouraging prototype.
On the side of SFD,  
  more attention should be paid to cooperativity. 
In particular, 
  it would be interesting 
  to devise an extension of SFD 
  in which $\alpha$ relaxation occurs cooperatively, 
  as it would shed more light on studies of glassy dynamics 
  than the classical SFD 
  which is essentially $\beta$ relaxation in 1D systems.

\section*{Acknowledgments}

We appreciate fruitful discussions 
  with Atsushi Ikeda, Takeshi Kawasaki, Makoto Iima, 
  Hajime Yoshino, Kunimasa Miyazaki,
  Akio Nakahara, and So Kitsunezaki. 
We also thank Hayato Shiba
  for drawing our attention to Ref.~\refcite{Jancovici.PRL19}
  and acknowledge linguistic advice from Rhodri Nelson.
The first author (Ooshida) 
  is grateful to the participants 
  in the Majorana Centre Conference on SFD, July 2014,  
  including---but not limited to---%
  Ophir Flomenbom, Alessandro Taloni,
  Eli Barkai, Michael~A.~Lomholt, 
  and Christophe Coste.
This work was supported 
  by Grants-in-Aid for Scientific Research 
  (\textsc{Kakenhi})  
  No.~21540388, No.~24540404, and No.~15K05213, 
  JSPS (Japan).


\begingroup 
  
\endgroup 


\small

\begin{thebibliography}{10}

\bibitem{Taloni.BRL9}
A.~Taloni and F.~Marchesoni, {\em Biophys. Rev. Lett.} {\bf 9}  (2014).

\bibitem{Coste.BRL9}
C.~Coste, J.-B. Delfau and M.~S. Jean, {\em Biophys. Rev. Lett.} {\bf 9}
  (2014).

\bibitem{Liu.Book2001}
A.~J. Liu and S.~R. Nagel, {\em Jamming and Rheology: Constrained Dynamics on
  Microscopic and Macroscopic Scales} (Taylor \& Francis, London, 2001).

\bibitem{Berthier.RMP83}
L.~Berthier and G.~Biroli, {\em Rev. Mod. Phys.} {\bf 83}, 587  (2011).

\bibitem{Das.Book2011}
S.~P. Das, {\em Statistical Physics of Liquids at Freezing and Beyond}
  (Cambridge Univ. Press, 2011).

\bibitem{Hoefling.RPP76}
F.~H{\"o}fling and T.~Franosch, {\em Rep. Prog. Phys.} {\bf 76},   046602
  (2013).

\bibitem{Yamamoto.PRE58}
R.~Yamamoto and A.~Onuki, {\em Phys. Rev. E} {\bf 58}, 3515  (1998).

\bibitem{Berthier.Book2011}
L.~Berthier, G.~Biroli, J.-P. Bouchaud, L.~Cipelletti and W.~van Saarloos
  (eds.), {\em Dynamical Heterogeneities in Glasses, Colloids, and Granular
  Media} (Oxford University Press, Oxford, 2011).

\bibitem{Goetze.TTSP24}
W.~G{\"o}tze and L.~Sj{\"o}gren, {\em Transport Theory and Statistical Physics}
  {\bf 6-8}, 801  (1995).

\bibitem{Kob.PRE51}
W.~Kob and H.~C. Andersen, {\em Phys. Rev. E} {\bf 51}, 4626  (1995).

\bibitem{Kob.PRE52}
W.~Kob and H.~C. Andersen, {\em Phys. Rev. E} {\bf 52}, 4134  (1995).

\bibitem{Reichman.JStat2005}
D.~R. Reichman and P.~Charbonneau, {\em J. Stat. Mech.} ,   P05013  (2005).

\bibitem{Otsuki.PRE86}
M.~Otsuki and H.~Hayakawa, {\em Phys. Rev. E} {\bf 86},   031505  (2012).

\bibitem{Doliwa.PRE61}
B.~Doliwa and A.~Heuer, {\em Phys. Rev. E} {\bf 61}, 6898  (2000).

\bibitem{Brito.JCP131}
C.~Brito and M.~Wyart, {\em J. Chem. Phys.} {\bf 131},   024504  (2009).

\bibitem{Sota.JKPS54}
S.~Sota and M.~Itoh, {\em Journal of Korean Physical Society} {\bf 54}, 386
  (2009).

\bibitem{Donati.PRL80}
C.~Donati, J.~F. Douglas, W.~Kob, S.~J. Plimpton, P.~H. Poole and S.~C.
  Glotzer, {\em Phys. Rev. Lett.} {\bf 80}, 2338  (1998).

\bibitem{Dasgupta.EPL15}
C.~Dasgupta, A.~V. Indrani, S.~Ramaswamy and M.~K. Phani, {\em Europhys. Lett.}
  {\bf 15}, 307  (1991).

\bibitem{Glotzer.JCP112}
S.~C. Glotzer, V.~N. Novikov and T.~B. Schr{\o}der, {\em J. Chem. Phys.} {\bf
  112}, 509  (2000).

\bibitem{Shiba.PRE86}
H.~Shiba, T.~Kawasaki and A.~Onuki, {\em Phys. Rev. E} {\bf 86},   041504
  (2012).

\bibitem{Kawasaki.PRE87}
T.~Kawasaki and A.~Onuki, {\em Phys. Rev. E} {\bf 87},   012312  (2013).

\bibitem{Ooshida.JPSJ80}
{{Ooshida}~T.}, S.~Goto, T.~Matsumoto, A.~Nakahara and M.~Otsuki, {\em J.
  Phys. Soc. Japan} {\bf 80},   074007  (2011).

\bibitem{Ooshida.PRE88}
{{Ooshida}~T.}, S.~Goto, T.~Matsumoto, A.~Nakahara and M.~Otsuki, {\em
  Phys. Rev. E} {\bf 88},   062108  (2013).

\bibitem{Majumdar.PhysicaA177}
S.~N. Majumdar and M.~Barma, {\em Physica A} {\bf 177}, 366  (1991).

\bibitem{Rallison.JFM186}
J.~M. Rallison, {\em J. Fluid Mech.} {\bf 186}, 471  (1988).

\bibitem{van-Beijeren.PRB28}
H.~van Beijeren, K.~W. Kehr and R.~Kutner, {\em Phys. Rev. B} {\bf 28}, 5711
  (1983).

\bibitem{Nelissen.EPL80}
K.~Nelissen, V.~Misko and F.~Peeters, {\em Europhys. Lett.} {\bf 80},   56004
  (2007).

\bibitem{Tkachenko.PRE82}
D.~Tkachenko, V.~Misko and F.~Peeters, {\em Phys. Rev. E} {\bf 82},   051102
  (2010).

\bibitem{Alexander.PRB18}
S.~Alexander and P.~Pincus, {\em Phys. Rev. B} {\bf 18},   2011  (1978).

\bibitem{Hodgkin.JPhysiol128}
A.~L. Hodgkin and R.~D. Keynes, {\em J. Physiol.} {\bf 128}, 61  (1955).

\bibitem{Harris.JAP2}
T.~E. Harris, {\em J. Appl. Probab.} {\bf 2}, 323  (1965).

\bibitem{Lefevre.PRE72}
A.~Lef{\`e}vre, L.~Berthier and R.~Stinchcombe, {\em Phys. Rev. E} {\bf 72},
  010301(R)  (2005).

\bibitem{Abel.PNAS106}
S.~M. Abel, Y.-L.~S. Tse and H.~C. Andersen, {\em Proc. Natl. Acad. Sci. USA}
  {\bf 106}, 15142  (2009).

\bibitem{Frusawa.PhysLettA378}
H.~Frusawa, {\em Phys. Lett. A} {\bf 378}, 1780  (2014).

\bibitem{Miyazaki.JCP117}
K.~Miyazaki and A.~Yethiraj, {\em J. Chem. Phys.} {\bf 117},   10448  (2002).

\bibitem{Pal.PRE78}
P.~Pal, C.~S. O'Hern, J.~Blawzdziewicz, E.~R. Dufresne and R.~Stinchcombe, {\em
  Phys. Rev. E} {\bf 78},   011111  (2008).

\bibitem{Hahn.JPhA28}
K.~Hahn and J.~K{\"a}rger, {\em J. Phys. A: Math. Gen.} {\bf 28}, 3061  (1995).

\bibitem{Kollmann.PRL90}
M.~Kollmann, {\em Phys. Rev. Lett.} {\bf 90},   180602  (2003).

\bibitem{Naegele.PhysRep272}
G.~N{\"a}gele, {\em Physics Reports} {\bf 272}, 215  (1996).

\bibitem{Dhont.Book1996}
J.~K.~G. Dhont, {\em An Introduction to Dynamics of Colloids} (Elsevier,
  Amsterdam, 1996).

\bibitem{Lucena.PRE85}
D.~Lucena, D.~Tkachenko, K.~Nelissen, Misko, Ferreira, Farias and Peeters, {\em
  Phys. Rev. E} {\bf 85},   031147  (2012).

\bibitem{Siems.SR2}
U.~Siems, C.~Kreuter, A.~Erbe, N.~Schwierz, S.~Sengupta, P.~Leiderer and
  P.~Nielaba, {\em Scientific Reports} {\bf 2},   1015  (2012).

\bibitem{Wanasundara.JCP140}
S.~N. Wanasundara, R.~J. Spiteri and R.~K. Bowles, {\em J. Chem. Phys.} {\bf
  140},   024505  (2014).

\bibitem{Ooshida@Rhodes1407+}
{{Ooshida}~T.}, S.~Goto, T.~Matsumoto and M.~Otsuki, Displacement
  correlation as an indicator of collective motion in one-dimensional and
  quasi-one-dimensional systems of repulsive {Brownian} particles,
  to be published in Mod. Phys. Lett. B.

\bibitem{Flomenbom.EPL83}
O.~Flomenbom and A.~Taloni, {\em Europhys. Lett.} {\bf 83},   20004  (2008).

\bibitem{Flomenbom.PRE82}
O.~Flomenbom, {\em Phys. Rev. E} {\bf 82},   031126  (2010).

\bibitem{Flomenbom.PLA374}
O.~Flomenbom, {\em Phys. Lett. A} {\bf 374}, 4331  (2010).

\bibitem{Flomenbom.EPL94}
O.~Flomenbom, {\em Europhys. Lett.} {\bf 94},   58001  (2011).

\bibitem{Ussing.ActaPhysiolScand19}
H.~H. Ussing, {\em Acta Physiol. Scand.} {\bf 19}, 43  (1949).

\bibitem{Hsieh.BiophysChem139}
C.-P. Hsieh, {\em Biophysical Chemistry} {\bf 139}, 57  (2009).

\bibitem{Hille.NM5}
B.~Hille, C.~M. Armstrong and R.~Mackinnon, {\em Nature Medicine} {\bf 5}, 1105
   (1999).

\bibitem{Hill.PNAS68.II}
T.~L. Hill and Y.-D. Chen, {\em Proc. Nat. Acad. Sci. USA} {\bf 68}, 1711
  (1971).

\bibitem{Barkai.PRL102}
E.~Barkai and R.~Silbey, {\em Phys. Rev. Lett.} {\bf 102},   050602  (2009).

\bibitem{Barkai.PRE81}
E.~Barkai and R.~Silbey, {\em Phys. Rev. E} {\bf 81},   041129  (2010).

\bibitem{Leibovich.PRE88}
N.~Leibovich and E.~Barkai, {\em Phys. Rev. E} {\bf 88},   032107  (2013).

\bibitem{Pusey.PRL59}
P.~N. Pusey and W.~van Megen, {\em Phys. Rev. Lett.} {\bf 59},   2083  (1987).

\bibitem{Taloni.PRE78}
A.~Taloni and M.~A. Lomholt, {\em Phys. Rev. E} {\bf 78},   051116  (2008).

\bibitem{Kutner.PRB30}
R.~Kutner, H.~van Beijeren and K.~W. Kehr, {\em Phys. Rev. B} {\bf 30}, 4382
  (1984).

\bibitem{van-Beijeren.PRL55}
H.~van Beijeren and R.~Kutner, {\em Phys. Rev. Lett.} {\bf 55}, 238  (1985).

\bibitem{Sellitto.PRE62}
M.~Sellitto and J.~J. Arenzon, {\em Phys. Rev. E} {\bf 62}, 7793  (2000).

\bibitem{Hansen.Book2006}
J.-P. Hansen and I.~R. McDonald, {\em Theory of Simple Liquids}, 3rd edn.
  (Academic Press, Amsterdam, 2006).

\bibitem{Dean.JPhAMG29}
D.~S. Dean, {\em J. Phys. A: Math. Gen.} {\bf 29},   L613  (1996).

\bibitem{Kawasaki.PhysicaA208}
K.~Kawasaki, {\em Physica A} {\bf 208}, 35  (1994).

\bibitem{Kawasaki.JStatPhys93}
K.~Kawasaki, {\em Journal of Statistical Physics} {\bf 93}, 527  (1998).

\bibitem{Hagen.PRL78}
M.~H.~J. Hagen, I.~Pagonabarraga, C.~P. Lowe and D.~Frenkel, {\em Phys. Rev.
  Lett.} {\bf 78}, 3785  (1997).

\bibitem{Alder.PRA1}
B.~J. Alder and T.~E. Wainwright, {\em Phys. Rev. A} {\bf 1}, 18  (1970).

\bibitem{Lizana.PRE81}
L.~Lizana, T.~Ambj{\"o}rnsson, A.~Taloni, E.~Barkai and M.~A. Lomholt, {\em
  Phys. Rev. E} {\bf 81},   051118  (2010).

\bibitem{Gotze.RPP55}
W.~G{\"o}tze and L.~Sj{\"o}gren, {\em Rep. Prog. Phys.} {\bf 55},   241
  (1992).

\bibitem{Goetze.Book2009}
W.~G{\"o}tze, {\em Complex Dynamics of Glass-Forming Liquids: A Mode-coupling
  theory} (Oxford University Press, New York, 2009).

\bibitem{Fedders.PRB17}
P.~A. Fedders, {\em Phys. Rev. B} {\bf 17}, 40  (1978).

\bibitem{Centres.PRE81}
P.~M. Centres and S.~Bustingorry, {\em Phys. Rev. E} {\bf 81},   061101
  (2010).

\bibitem{Doi.Book1986}
M.~Doi and S.~F. Edwards, {\em The Theory of Polymer Dynamics} (Oxford, 1986).

\bibitem{Majumdar.PRB44}
S.~N. Majumdar and M.~Barma, {\em Phys. Rev. B} {\bf 44},   5306  (1991).

\bibitem{Edwards.PRSLA381}
S.~F. Edwards and D.~R. Wilkinson, {\em Proc. R. Soc. London, Ser. A} {\bf
  381}, 17  (1982).

\bibitem{Krug.AdvPhys46}
J.~Krug, {\em Adv. in Phys.} {\bf 46}, 139  (1997).

\bibitem{Brito.EPL76}
C.~Brito and M.~Wyart, {\em Europhys. Lett.} {\bf 76}, 149  (2006).

\bibitem{Toninelli.PRE71}
C.~Toninelli, M.~Wyart, L.~Berthier, G.~Biroli and J.-P. Bouchaud, {\em Phys.
  Rev. E} {\bf 71},   041505  (2005).

\bibitem{Jancovici.PRL19}
B.~Jancovici, {\em Phys. Rev. Lett.} {\bf 19}, 20  (1967).

\bibitem{Diamant.arXiv1406}
H.~Diamant, Criteria of amorphous solidification, {arXiv:1406.2508}.

\bibitem{Kraichnan.PhF8}
R.~H. Kraichnan, {\em Physics of Fluids} {\bf 8}, 575  (1965).

\bibitem{Kaneda.JFM107}
Y.~Kaneda, {\em J. Fluid Mech.} {\bf 107}, 131  (1981).

\bibitem{Kida.JFM345}
S.~Kida and S.~Goto, {\em J. Fluid Mech.} {\bf 345}, 307  (1997).

\bibitem{Bird.Book1987}
R.~B. Bird, R.~C. Armstrong and O.~Hassager, {\em Dynamics of polymeric
  liquids}, second edn. (Wiley, New York, 1987).

\bibitem{Kraichnan.JFM5}
R.~H. Kraichnan, {\em J. Fluid Mech.} {\bf 5}, 497  (1959).

\bibitem{Goto.PhysicaD117}
S.~Goto and S.~Kida, {\em Physica D} {\bf 117}, 191  (1998).

\bibitem{Kob.PRL78}
W.~Kob and J.-L. Barrat, {\em Phys. Rev. Lett.} {\bf 78}, 4581  (1997).

\bibitem{collab3.RR@PRE}
{{Ooshida}~T.}, S.~Goto, T.~Matsumoto and M.~Otsuki, Calculation of
  displacement correlation tensor indicating vortical cooperative motion in
  two-dimensional colloidal liquids (in preparation).

\bibitem{Adam.JCP43}
G.~Adam and J.~H. Gibbs, {\em J. Chem. Phys.} {\bf 43}, 139  (1965).

\bibitem{Kawasaki.JPhys23}
T.~Kawasaki and H.~Tanaka, {\em J. Phys.: Condens. Matter} {\bf 23},   194121
  (2011).

\end{thebibliography}
\end{document}